\documentclass[ascmo, manuscript]{copernicus}

\begin{document}

\begin{nolinenumbers}

\title{Post-processing of wind gusts from COSMO-REA6 with a spatial Bayesian hierarchical extreme value model\footnote{This manuscript has been submitted to Advances in Statistical Climatology, Meteorology and Oceanography (ASCMO) and is currently under review.}}


\Author[][pertz@uni-bonn.de]{Philipp}{Ertz} 
\Author[]{Petra}{Friederichs}

\affil[]{Institute of Geosciences, University of Bonn, Auf dem H\"{u}gel 20, 53121 Bonn, Germany}




\runningtitle{Spatial Bayesian modeling of wind-gust extremes}

\runningauthor{Ertz \& Friederichs}

\received{}
\pubdiscuss{} 
\revised{}
\accepted{}
\published{}


\firstpage{1}

\maketitle

\begin{abstract}
The aim of this study is to provide a probabilistic gust analysis for the region of Germany that is calibrated with station observations and with an interpolation to unobserved locations. To this end, we develop a spatial Bayesian hierarchical model (BHM) for the post-processing of surface maximum wind gusts from the COSMO-REA6 reanalysis. Our approach uses a non-stationary extreme value distribution for the gust observations at the top level, with parameters that vary according to a linear model using COSMO-REA6 predictor variables. To capture spatial patterns in surface extreme wind gust behavior, the regression coefficients are modeled as 2-dimensional Gaussian random fields with a constant mean and an isotropic covariance function that depends only on the distance between locations. In addition, we include an elevation offset in the distance metric for the covariance function to account for differences in topography. This allows us to include data from mountaintop stations in the training process and to utilize all available information. The training of the BHM is carried out with an independent data set from which the data at the station to be predicted are excluded. We evaluate the spatial prediction performance at the withheld station using Brier score and quantile score, including their decomposition, and compare the performance of our BHM to climatological forecasts and a non-hierarchical, spatially constant baseline model. This is done for 109 weather stations in Germany. Compared to the spatially constant baseline model, the spatial BHM significantly improves the estimation of local gust parameters. It shows up to 5 \unit{\%} higher skill for prediction quantiles and provides a particularly improved skill for extreme wind gusts. In addition, the BHM improves the prediction of threshold levels at most of the stations. Although a spatially constant approach already provides high skill, our BHM further improves predictions and improves spatial consistency. 
\end{abstract}


\introduction  

Wind warnings are among the most frequent types of weather warning issued by the German Meteorological Service (DWD).
Key users are emergency managers, air and rail traffic, energy companies, and the general public. 
Improving the quality of extreme wind warnings is therefore an important task for, e.g., meteorological services such as the DWD.  
A central characteristic of atmospheric wind speed is its high spatial variability with a comparatively sparse observation network for wind speed measurements. In addition, severe wind gusts are recurring but rare, so they are inherently extreme events. This makes forecasting a challenge and complicates the verification of wind gust warnings.
A suitable representation of wind gusts that includes uncertainties in reanalysis models is therefore essential for the development of an effective warning strategy.

State-of-the-art numerical weather prediction (NWP) models provide a deterministic wind gust diagnostic based on the current simulated turbulent and convective state \citep[e.g.][]{Schulz03, Schulz08}. 
\citet{Brasseur01} proposed a peak wind speed diagnostic, which assumes that gusts are generated by the deflection of high-flowing air parcels in the upper boundary layer and are brought down by turbulent eddies. 
The deflection of high-flowing parcels to the surface occurs when the vertical component of the turbulent kinetic energy (TKE) associated with a given air parcel is strong enough to overcome the buoyancy forces. 
Another mechanism for the generation of surface wind gusts is described by \citet{Nakamura96}. 
They suggested that the gusts originate from a convective downdraft that is deflected horizontally when it reaches the surface. 
Both theories motivate a diagnosis in NWP models for wind gust forecasts based on turbulent kinetic energy and the stability of the vertical stratification of the atmosphere. However, these are mostly deterministic in nature and do not incorporate uncertainty in the gust forecasts.

One approach for addressing uncertainties in the description of wind gusts is a probabilistic post-processing providing a conditional probability distribution for surface wind gusts based on predictor variables from the reanalysis. Given that wind gusts represent extreme values, extreme value statistics is a suitable probabilistic model \citep{Coles01, Beirlant04}.
A basic application of extreme value statistics to maximum wind gusts can be found in \citet{Walshaw94}, who develop a threshold model for wind gusts. 
The model is expanded with spatial covariates by \citet{Walshaw00}. They demonstrate that the model performance at single locations can be enhanced by the use of predictor variables such as the local mean wind speed.
These simple models can be readily adapted into a post-processing approach by taking time series of the covariates for the model parameters. As shown in \citet{Friederichs09} and \citet{Friederichs18}, these simple approaches already possess high skill.

\citet{Wessel25} present an interesting approach for the post-processing of extreme wind speed, based on ensemble model output statistics \citep{Gneiting05}. They do not use extreme value theory but train truncated normal and logistic distributions using threshold weighted scoring rules to enhance the predictive performance for extremes. To increase the training data size, data from stations with similar wind characteristics, based on a clustering approach, is pooled together and one model is trained for each cluster of stations. However, it is also possible to include a spatial structure directly into the model by modeling the dependency structure between stations. 

Modeling spatial dependency in extreme value theory poses the challenge that there are no parametric forms for multivariate extreme value distributions \citep{Coles01}.
\citet{Davison12} present several approaches to incorporate the spatial dependency structure into statistical extreme value models. These approaches include latent spatial process variables \citep{Cooley07, Apputhurai13, Stephenson16},  copula models \citep{Sang10} and max-stable processes \citep{Oesting16, Davison12b}.
Latent process variables are introduced in a Bayesian hierarchical model (BHM) framework, while assuming conditional independence between gusts at different locations, whereas copula models and max-stable processes explicitly model dependencies between gusts at different locations.
\citet{Davison12} demonstrate that latent process variables offer the greatest flexibility, while possessing a high degree of accuracy. However, the assumptions on stationarity and the covariance function may result in the generation of implausible extremal patterns.

The general theory for hierarchical spatial process models is outlined, for example, in \citet{Banerjee03}.
The use of a BHM allows for the pooling of information across locations and time, facilitating an optimal spatial and temporal representation.
A groundbreaking example of a latent Gaussian process model for extreme weather can be found in \citet{Cooley07}.
The authors construct a hierarchical Bayesian extreme value model for maximum precipitation in Colorado. At the initial level, they utilize a generalized Pareto distribution whose parameters vary as a function of covariates and a latent spatial Gaussian process in climate space, which is defined by elevation and mean precipitation.
More recent examples of spatial modeling of extreme values in atmospheric sciences include the model developed by \citet{Apputhurai13}, which was used to predict extreme precipitation in Western Australia. This model employed a spatio-temporal hierarchical model with anisotropic Gaussian random fields (GRFs) based on spatial explanatory variables and included trends in the model parameters. 
\citet{Dyrrdal15} model extreme precipitation return levels in Norway in a similar vein.

In \citet{Friederichs18}, the authors fit a regression model to the predictor variables of the NWP model using a conditional Gumbel distribution for the marginals. 
Subsequently, the non-explained residuals are modeled as a bivariate Brown-Resnick process to account for the remaining spatial dependency structure.
Similarly, \citet{Oesting16} use a bivariate, maximally stable Brown-Resnick process and Gumbel margins to directly model the dependency between forecasts and observed wind gust fields. 

\citet{Baran24} demonstrate, with reference to the example of 10 m wind speed, that the spatial interpolation of post-processing schemes can achieve a high level of skill at locations where observations are lacking.
The interpolated model demonstrates superior skill compared to both a globally trained model and a locally trained model at each individual location. 
Consequently, it can be concluded that the spatial interpolation of post-processing approaches effectively addresses the challenge of limited training data for post-processing gridded data sets, due to the restricted number and irregular distribution of observation stations.

This paper presents an integration of the post-processing approach originally proposed by \citet{Friederichs18} with the spatial hierarchical modeling techniques derived from the precipitation model developed by \citet{Cooley07}.
We utilize the flexible implementation of latent process models, assuming conditional independence between maxima at two different locations.
As wind gusts are typically very localized phenomena, the spatial dependency of maxima at two locations can be disregarded during modeling. Furthermore, our model is designed for spatial interpolation and incorporates spatio-temporal predictors for the generalized extreme value distribution (GEV) parameters, enabling us to interpolate the model to unobserved locations via kriging \citep{Stein99}. This provides comprehensive post-processing across the entire NWP model domain.

Our methodology is similar to the spatially adaptive Bayesian post-processing approach by \citet{Moeller16} for ensemble temperature forecasts.
It employs Gaussian Markov random fields and integrated nested Laplace approximations (INLA) of stochastic partial differential equations (SPDE) \citep{Rue09, Lindgren11}.
However, as we investigate wind gusts, we are examining non-Gaussian distributed observations, and in addition, we are incorporating a spatially variable and non-stationary variance.
Furthermore, our post-processing approach is currently not designed for ensemble predictions, but it can be readily extended to ensemble post-processing through the selection of suitable predictors, such as the ensemble mean.

The remainder of this paper is structured as follows. Sect.~\ref{sec:Data} presents the data used in this study and in Sect.~\ref{sec:Methods}, we present our model formulation. The methods used for training and evaluating the model are explained in Sect.~\ref{sec:Training}. Sect.~\ref{sec:Results} will show and discuss the results of the verification on observational data. Finally, we will conclude with a summary of our findings in Sect.~\ref{sec:Conclusions}.

\section{Data}\label{sec:Data}

\subsection{Station data}
The wind gust observations $\mathrm{FX}$ were selected from a total of 109 synoptic observation stations (SYNOP), including mountain top stations.
Please refer to Fig.~\ref{fig:SkillLocMod} in Sect.~\ref{sec:ConstMod} for a map of the corresponding locations. Additionally, Fig.~\ref{fig:SkillLocMod} provides an overview over mean wind characteristics at these locations.
Data are retrieved from the DWD Climate Data Center \citep{cdc}. DWD performs basic quality control of the incoming data.
The stations are selected based on the least number of missing observations between 1995 and 2018.
The observation time series are available in hourly resolution.

The two fundamental mechanisms of wind gusts production are shear-driven turbulent wind gusts and convective wind gusts driven by thunderstorm activity \citep{Bradbury94}.
Given the increased convective activity during the summer months, it is expected that the behavior of wind gust will differ from that observed in winter, when the characteristics of gusts are largely influenced by the occurrence of winter storms.
Accordingly, the data set employed in this study comprises observations from the warm months between May and October, thus avoiding the need to consider the annual cycle of deep convection. 
Similarly, wind gusts display a pronounced diurnal cycle. During nocturnal hours, a stable boundary layer forms and lower wind gust values are observed, whereas stronger peak gusts occur during the convective maximum in the late afternoon. To include the peak gusts of the day, our investigation window is focused on the afternoon hours from 13--18 UTC. Then, we generate daily time series using the maximum value inside the window for each day.

The availability of wind gust measurements from the years preceding 2001 is limited, and thus these years have been excluded from the analysis. Additionally, to ensure spatial homogeneity within the data set, days with fewer than four measurements within the specified investigation window at a given station have been omitted. 

\subsection{COSMO-REA6}
COSMO-REA6 \citep{Bollmeyer15} is a regional reanalysis model for the CORDEX-EURO11 domain \citep{Giorgi09} based on the COnsortium for Small scale MOdelling (COSMO) climate model COSMO-CLM \citep{Rockel08}. 
It was developed through a collaborative effort between the Hans-Ertel-Center for Weather Research (HErZ), Climate Monitoring and Diagnostics, from Bonn and Cologne Universities and DWD.
The horizontal grid spacing is $0.055^{\circ}$, which is approximately 6 $\unit{km}$, and the model operates on 40 vertical levels. The data included in this study encompass the 10 $\unit{m}$ values for mean horizontal wind, obtained from the horizontal components U\_10M and V\_10M, and the 10 $\unit{m}$ horizontal maximum wind speed VMAX\_10M. The deterministic gust diagnostic in COSMO-CLM, VABSMX\_10M, is not publicly available in COSMO-REA6.
In the following sections, mean horizontal wind will be referred to as $V_{\mathrm{m}}$ and VMAX\_10M as $V_{\mathrm{max}}$.

The 2D fields from COSMO-REA6 are available in 15 minute intervals from 1995 to 2019. 
However, we retrieved hourly values from COSMO-REA6 between 12--18 UTC from May to October 2001 to 2018, to be in line with the SYNOP observations. 
The reanalysis time series were aligned with the weather stations by selecting the closest grid cell to the station in terms of distance on a sphere (distance along great circles). 
Subsequently, the reanalysis data underwent the same preprocessing as the observational data, with daily maxima employed for $V_{\mathrm{max}}$ and daily mean values for $V_{\mathrm{m}}$.

For the purpose of our simulation study, we assume that the predictor data from COSMO-REA6 and the wind gust observations are independent. This might not necessarily be the case, as mean wind observations from the SYNOP stations are assimilated into COSMO-REA6 and usually, wind gust and mean wind observations are collinear to some degree. However, as the gust observations are not assimilated directly and we do not use any other observational data, the assumption of independence should hold.
However, it should be noted that the spatial alignment of weather stations and model grid cells is also imperfect, and that a statistical post-processing is precisely used to address model deficiencies in the reanalysis.

\section{Model formulation}\label{sec:Methods}

We propose a Bayesian hierarchical model based on the extreme precipitation model of \citet{Cooley07} for the spatial post-processing of wind gusts. 
The complete version of the spatial Bayesian hierarchical model, henceforth referred to as SpatBHM, is depicted as a directed acyclic graph in Fig.~\ref{fig:DAGSpatbhm}.
For readers who are unfamiliar with Bayesian hierarchical models, we recommend consulting \citet[Chapter 5]{Gelman14}.

Our model comprises three levels: a data level, which describes the actual modeling of wind gust observations using extreme value statistics; an underlying predictor level which includes the predictor variables from COSMO-REA6; and a prior level, which contains our prior knowledge about the model hyperparameters (cf. Fig.~\ref{fig:DAGSpatbhm}), as required for Bayesian inference.
To accommodate the spatial dependency structure, we introduce an additional level between the predictor level and the prior level.
This level will henceforth be referred to as the spatial dependency level or process level.

\begin{figure}
    \centering
    \includegraphics[width=\textwidth]{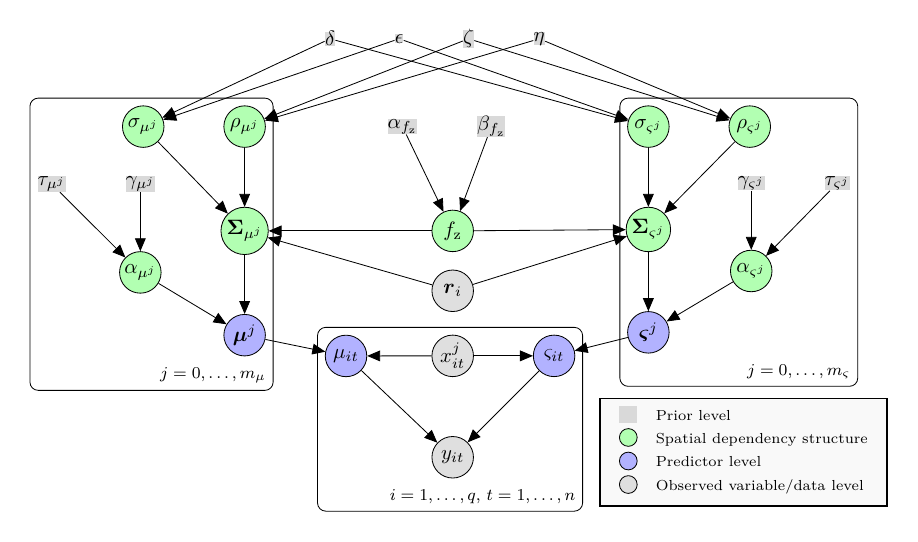}
    \caption{Directed acyclic graph for \mbox{SpatBHM} with predictand $y_{ik}$, covariates $x_{j,ik}$ and station coordinates $\vec{r}_i$, for location $i$ and time $t$. $m_{\mu/\varsigma}$ refer to the respective number of covariates for location and scale. Please refer to Table~\ref{tab:notation} for a comprehensive notation reference.}
    \label{fig:DAGSpatbhm}
\end{figure}

\subsection{Data level}\label{Sec:DataLevel}
Extreme value theory provides an asymptotic theory for the distribution for sample maxima \citep{Coles01, Beirlant04}.
As gust measurements refer to the highest 3 s average wind speed  recorded over the course of a 1 h interval, a generalized extreme value (GEV) distribution  serves as a theoretically consistent model for gust observations.
The GEV distribution is characterized by three parameters: location $\mu$, scale $\varsigma$ and shape $\xi$.
These parameters may vary both spatially and temporally.
However, allowing for non-stationarity in $\xi$ largely increases the sampling  uncertainty not only for $\xi$, but also for the other parameters \citep{Friederichs09}. 
Therefore we decided to use a constant shape parameter.
To obtain an estimate of a reasonable $\xi$-value, we conducted single station fits for all weather stations. These fits showed either no clear signal or resulted in slightly negative values, implying a Weibull-type GEV (Appendix  \ref{A:GEV_shape}). The Weibull-type GEV imposes an unrealistic upper bound for wind gusts extremes \citep{Perrin06}.
In order to obtain stable parameter estimates and avoid making assumptions regarding an upper bound for wind gusts, we assume $\xi=0$, resulting in a Gumbel distribution. 
We model the gust measurements $y_{it}$ at location $\vec{r}_i$, $i \in \{1, \ldots, q\}$, and time $t\in \{1, \ldots, n\}$ as a Gumbel-distribution
\begin{equation}\label{eq:Gumbel}
    G(y_{it}) = \exp\left(-\exp\left(-\frac{y_{it}-\mu_{it}}{\varsigma_{it}}\right)\right) 
\end{equation}
with non-stationary location $\mu_{it}$ ans scale $\varsigma_{it}$.
As indicated by the indices $i = 1,\ldots, q$ for location and $t = 1,\ldots, N$ for time, the Gumbel parameters are allowed to vary in space and time. The reader is referred to Table~\ref{tab:notation} in Appendix~\ref{A:symbols} for a thorough list of notations.

\subsection{Predictor level}\label{sec:PredictorLevel}

On the next underlying level, the predictor level, we introduce spatio-temporal predictor variables $x_{j,ik}$ as covariates for $\mu_{it}$ and $\varsigma_{it}$.
The Gumbel parameters $\mu_{it}$ and $\varsigma_{it}$ are constructed as a linear model with regression coefficients $\mu^j_i$ and $\varsigma^j_i$
for the $j$th predictor variable.
The vector of the predictor variables is constructed including a constant intercept, so that $\vec{x}^\mu_{it} = (1, x_{1,it}, \ldots, x_{m_\mu,it})$ and $\vec{x}^\varsigma_{it} = (1, x_{1,it}, \ldots, x_{m_\varsigma,it})$.
With $\vec{\beta}_i^{\mu} = (\mu^{0}_{i}, \ldots, \mu^{{m_\mu}}_{i})^T$ and $\vec{\beta}_i^{\varsigma} = (\varsigma^{0}_{i}, \ldots, \varsigma^{{m_\varsigma}}_{i})^T$, we obtain 
\begin{eqnarray}\label{eq:regressions}
    \mu_{it} &=&  \vec{x}^{\mu}_{it}\boldsymbol{\beta}^{\mu}_i\notag\\
    \varsigma_{it} &=& \exp( \vec{x}^{\varsigma}_{it}\boldsymbol{\beta}^{\varsigma}_i).
\end{eqnarray}
An exponential link for $\varsigma_{it}$ ensures positive definiteness of the scale parameter. 
The regression coefficient vectors $\vec{\beta}_i^\mu$ and $\vec{\beta}_i^\varsigma$ are stationary over time, but vary in space.

\subsection{Spatial dependency structure}
To address the spatial dependency structure, we assume that the regression coefficients $\mu^j_i$ and $\varsigma^j_i$ are realizations of a spatial random process observed at locations $\vec{r}_i$, $i = 1, \ldots, q$. These spatial random processes will be referenced by $\mu^{j}(\vec{r})$ and $\varsigma^{j}(\vec{r})$, respectively, where $j$ indicates the respective covariate.
We therefore assume that the regression coefficients, which in turn determine the GEV parameters, are spatially dependent fields. This approach enables the integration of data within the spatial domain, thereby facilitating the incorporation of information regarding the distribution at neighboring stations. However, the degree of the spatial dependency is estimated in the BHM. In order to model the spatial dependency structure, a Gaussian process level is introduced below the predictor level. This means that each spatial regression coefficient is assigned a spatial process in the form of a GRF, namely $\mu^{j}(\vec{r})$ and $\varsigma^{j}(\vec{r})$, where $\vec{r} $ represents the spatial coordinates and $\mu^j_i = \mu^j(\vec{r}_i)$ and $\varsigma^j_i = \varsigma^j(\vec{r}_i)$.

If $\mu^{j}(\vec{r})$ is a GRF, then $\vec{\mu}^{j} = (\mu^{j}(\vec{r}_1), \ldots, \mu^{j}(\vec{r}_q))^T $  is multivariate normally distributed random vector. 
This applies analogously to $\vec{\varsigma}^{j} = (\varsigma^{j}(\vec{r}_1), \ldots, \varsigma^{j}(\vec{r}_q))^T$.
We assume the GRFs as spatially homogeneous with constant expectation values $\alpha_{\mu^{j}}(\vec{r})=\alpha_{\mu^{j}}$ and $\alpha_{\varsigma^{j}}(\vec{r})=\alpha_{\varsigma^{j}}$, respectively.
The process level for the regression coefficients then reads
\begin{eqnarray}\label{qq:SpatBHMloc}
\vec{\mu}^j &\sim  & {\cal MVN} ( \alpha_{\mu^{j}}\vec{I}, \boldsymbol{\mathrm{\Sigma}}_{\mu^{j}} )
\end{eqnarray}
and 
\begin{eqnarray}\label{Eq:SpatBHMscale}
\vec{\varsigma}^j &\sim  & {\cal MVN} ( \alpha_{\varsigma^{j}}\vec{I},\boldsymbol{\mathrm{\Sigma}}_{\varsigma^{j}} ).
\end{eqnarray}
Here, $\vec{I}$ is a unit vector of length $q$.
We further assume spatially homogeneous and isotropic covariance functions such that $\operatorname{Cov}(\mu^{j}(\vec{r}_i),\mu^{j}(\vec{r}_k)) = C_{\mu^{j}}(h_{ik})$ and $\operatorname{Cov}(\varsigma^{j}(\vec{r}_i),\varsigma^{j}(\vec{r}_k)) = C_{{\varsigma^{j}}}(h_{ik})$, where $h_{ik}=|\vec{r}_i-\vec{r}_k|$ is the distance between two locations.

In our study we use a Mat\'ern-class isotropic covariance function with parameters $\vec{\lambda}=(\sigma,\nu,\rho)$.
In general, the Mat\'ern covariance function takes a functional form involving the gamma function $\Gamma(x)$ and modified Bessel functions of the second kind $\mathcal{K}_\nu(x)$. 
In the case of half-integer values for the smoothness parameter $\nu$, this expression can be simplified as a product of a polynomial and an exponential term \citep{Rasmussen08, Gneiting12}. 
Therefore, we decided to fix the smoothness parameter at $\nu=3/2$ for numerical convenience.
With the definitions above for the parameters we obtain the following formulation for the covariance functions
\begin{eqnarray}\label{eq:maternLoc}
    \operatorname{Cov}_{\mu^{j}}(h) &= \left(\sigma_{\mu^{j}}\right)^2\left(  1 + \frac{\sqrt{3}h}{\rho_{\mu^{j}}}\right) \exp{\Bigl(-\frac{\sqrt{3}h}{\rho_{\mu^{j}}} \Bigr)}
\end{eqnarray}
and 
\begin{eqnarray}\label{eq:maternSig}
    \operatorname{Cov}_{\varsigma^{j}}(h) &= \left(\sigma_{\varsigma^{j}}\right)^2\left(  1 + \frac{\sqrt{3}h}{\rho_{\varsigma^{j}}}\right) \exp{\Bigl(-\frac{\sqrt{3}h}{\rho_{\varsigma^{j}}} \Bigr)} .
\end{eqnarray}
The Mat\'ern-class covariance models are recommended by \citet{Stein99} for the statistical interpolation of spatial data, due to their mean squares differentiability and their suitability for application in more than one dimension. 
Moreover, preliminary tests demonstrated a higher numerical stability for the Mat\'ern-model than for the exponential covariance function.

The introduction of the covariance function results in the incorporation of two additional parameters for each covariate at the process level.
In the following, the two parameters, $\sigma_{\mu^{j}/\varsigma^{j}}$ and $\rho_{\mu^{j}/\varsigma^{j}}$, are referred to as the sill and range of the covariance function for $\mu^{j}(\vec{r})$ and $\varsigma^{j}(\vec{r})$, although these terms do not directly correspond to the terms in the exponential covariance function.
Therefore, terms `process variance' and `correlation length' also exist \citep{Gneiting12}.
In contrast to the approach proposed by Stein (1999), we directly model the covariance function using the expression $(\sigma_{\mu^{j}})^2$ and $(\sigma_{\varsigma^{j}})^2$ (i.e.~in form of a variance), rather than the more conventional form of a standard deviation. 
This is done for the purpose of ensuring consistency in the variance of the GRF at a given location and the variance of the marginal normal distribution at that location.

\subsection{Distance metric for the spatial dependency structure}

We model the spatial fields on the Earth's surface in geographical coordinates with longitude $\lambda$, latitude $\varphi$ and altitude $z$, yielding $\vec{r} = (\lambda,\varphi,z)$. Therefore, the distance between the training locations is calculated using the distance on a sphere, often called great-circle distance, which we calculate 
as 
\begin{equation}\label{eq:GreatCircleDist}
    d_\mathrm{gc}(\vec{r}_i, \vec{r}_k) = 2R_\mathrm{E}\arcsin\sqrt{\frac{1-\cos{\Delta\varphi} + \cos\varphi_i\cos\varphi_k(1-\cos{\Delta\lambda)}}{2}},
\end{equation}
for $\Delta\varphi=\varphi_j-\varphi_i$ and $\Delta\lambda=\lambda_j-\lambda_i$ with the Earth's radius $R_\mathrm{E}$ that we take at 6371 km.
The inclusion of mountain top stations into the training data poses a challenge for the spatial interpolation due to the existence of valley locations with nearby mountain stations (e.g. Braunlage and Brocken in the Harz Mountains). Despite considerable differences in gust behavior due to the different topographical altitudes, the two stations are treated as if they were close to each other. This deteriorates the estimation of the dependency structure at all stations.
It is therefore necessary to achieve a separation of valley and mountain stations to improve the estimation of the spatial dependency structure.
To this end, the station altitude is included in the distance metric used for the covariance structure of the GRFs. This is achieved by introducing an elevation offset to the distance between mountain and valley locations.
The altitude difference between two stations is scaled by a factor $f_\mathrm{z}$ and added as an increment to the distance along a great circle yielding
\begin{equation}\label{eq:AltitudeScaling}
    d_\mathrm{v}(\vec{r}_i,\vec{r}_k) = f_\mathrm{z}(|z_k - z_i|),
\end{equation}
where $|\cdot|$ represents the absolute value, and $z_i$  the station altitude at location $\vec{r}_i$.  $f_z$ is the scaling factor, which we treat as an additional parameter to be estimated.
As the Mat\'ern-covariance function requires a Euclidean metric, the full distance between locations $i,k$ is calculated as
\begin{equation}\label{eq:AltitudeEuclidean}
    d(\vec{r}_i,\vec{r}_k) = \sqrt{d^2_\mathrm{gc}(\vec{r}_i,\vec{r}_k) + f_\mathrm{z}^2(z_k - z_i)^2}.
\end{equation}
This approach artificially increases the distance between two stations that are in close proximity but at markedly distinct elevations, while maintaining the general spatial dependency structure for locations in flat terrain.

As noted by \citet{Gneiting13}, Mat\'ern covariance functions are not positive definite on the sphere when combined with the great-circle distance and a smoothness parameter $\nu > 0.5$. Nevertheless, we adopted the Mat\'ern-3/2 covariance model because its local smoothness properties yielded numerically more stable covariance matrices for the short distances present in our data. In addition, since we only model a small portion of the sphere, the numerical distortions that could otherwise result in invalid covariance matrices, are unlikely to arise.

To verify that our covariance matrices remain positive definite in practice, we conducted an experiment using the station coordinates and the prior distributions described in Sect.~\ref{Sec:priors} As shown in Appendix \ref{A:CovMatrix}, this analysis revealed a parameter region in which the covariance matrices became non–positive definite, specifically when $\rho \sim \mathcal{O}(10^5)$ \unit{km}. However, as the posterior ranges in our application remain within $\mathcal{O}(10^3)$ \unit{km}, we conclude that the chosen combination of distance metric and covariance function is valid for the spatial domain under consideration.

\section{Model training and verification} \label{sec:Training}

\subsection{Specifying prior distributions}\label{Sec:priors}
\begin{table}[t]
    \caption{Prior distributions of the model parameters used for \mbox{SpatBHM}. The parameters of the spatial dependency structure, $\sigma$ and $\rho$, are simulated using the same prior distribution for all regression coefficients $\mu^{j}$ and $\varsigma^{j}$.}
    \label{tab:priors}
    \centering
    \begin{tabular}[t]{lr}
    \tophline
    \textbf{Parameter} & \textbf {Prior SpatBHM}\\
    \middlehline
    $\mu^z$ & $\mathcal{N}(\gamma_{\mu^z}, \tau_{\mu^z}^2)\equiv\mathcal{N}(0.23,0.1)$\\
    $f_\mathrm{z}$ & $\operatorname{Gamma}(15,0.15)$\\
    $\alpha_{\mu^{j}/\varsigma^{j}}$ & $\mathcal{N}(\gamma_{\mu^{j}/\varsigma^{j}}, \tau_{\mu^{j}/\varsigma^{j}}^2)$\\
    $\sigma_{\mu^{j}/\varsigma^{j}}$ & $\operatorname{InvGamma}(\delta=5,\epsilon=2)$\\
    $\rho_{\mu^{j}/\varsigma^{j}}$ & $\operatorname{InvGamma}(\zeta=1.05,\eta=100)$\\
    \bottomhline
    \end{tabular}
\end{table}

In order to perform Bayesian inference, it is necessary to specify prior distributions for the model parameters.
These prior distributions summarize the researcher's prior knowledge and assumptions regarding the distribution of these parameters. To ensure a fast convergence of the Markov chains, we use informative priors for the means of the GRFs, $\alpha_{\mu^{j}/\varsigma^{j}}$ and weakly informative priors for the parameters of the spatial dependency structure $\sigma_{\mu^{j}/\varsigma^{j}},\rho_{\mu^{j}/\varsigma^{j}}$. The impact of the priors in our configuration is small, as the extensive data set reduces the contribution of the prior belief on the inference process. Thus, the primary information is drawn from the training data. A summary of the prior distributions used in this study is presented in Table~\ref{tab:priors}.

For the expectation values $\alpha_{\mu^{j}/\varsigma^{j}}$ of the GRFs, we choose normal priors with parameters $\gamma_{\mu^{j}/\varsigma^{j}}$ and $\tau^2_{\mu^{j}/\varsigma^{j}}$. 
We first estimate the $\mu^{j}, \varsigma^{j}$ from the linear model individually at each station.
Subsequently, the prior distribution assumes the expectation values for each coefficient from the fit of the linear model on all stations simultaneously. The prior variances were informed by the variability of the results from individual station fits and rounded to facilitate interpretation and reporting. The resulting prior distributions for the model defined in Sect.~\ref{sec:ConstMod} are $\vec\mu\sim\mathcal{N}\left(\vec\gamma=(4.74,1.49,-0.89,0.22)^T,\vec\tau=(2,1,0.3,0.1)^T\right)$ and $\vec\varsigma\sim\mathcal{N}\left(\vec\gamma=(0.53,0.4,-0.23)^T,\vec\tau=(0.2,0.2,0.1)^T\right)$.

In the case of positive definite parameters, such as the sill and range of the GRFs, inverse gamma priors are selected.
We use the same  priors for all spatial dependence structures, independent of $\mu^{j}/\varsigma^{j}$. Consequently, a single prior value is set for the shapes $\delta, \zeta$ and the scales $\epsilon, \eta$ of the inverse gamma distributions.
All sill parameters are assigned $\sigma_{\mu^{j}/\varsigma^{j}} \sim \operatorname{InvGamma}(5,2)$, constraining them to a reasonable value range regarding the variability of the parameters. All range parameters are assigned $\rho_{\mu^{j}/\varsigma^{j}} \sim \operatorname{InvGamma}(1.05,100)$, favoring values between 0 and 500 km but otherwise not effecting major influence on the parameter estimation.

The scaling factor $f_z$ for the elevation offset in Eq.~(\ref{eq:AltitudeScaling}) is assigned a Gamma-distributed prior, to ensure positive definiteness. 
The expectation value of $f_z$ is derived from a quasigeostrophic scale analysis and is set to the ratio of the Coriolis force and the buoyancy force on synoptic scales.
This ratio is approximated as $\frac{N}{f}$ with the Brunt-V\"ais\"al\"a frequency $N$ and the Coriolis parameter $f$. 
In the mid-latitudes, typical values are found around 100. 
Hence, we model $f_z$ as $\operatorname{Gamma}(15,0.15)$.

\subsection{Parameter estimation}\label{sec:Mcmc}

The Bayesian parameter estimation was performed using Stan \citep{Stan}. 
Stan is a statistical software package designed for Bayesian inference based on Markov chain Monte Carlo (MCMC) techniques.
In MCMC, the posterior distribution of the parameters is approximated through the sampling of Markov chains, whose transition properties are tuned such that they converge to a target distribution \citep{Gelman14, Gilks95}.
We accessed Stan via its python interface, pystan \citep{Riddel21}.
Stan makes use of the No-U-Turn-sampler \citep[NUTS, ][]{Hoffman14}, which presents an adaptive and tuning-easy implementation of Hamiltonian Monte Carlo, leading to reasonably quick convergence of the Markov chains while remaining user-friendly.

For the training of SpatBHM, the sampling parameters for the tuning of the step size and the acceptance rate of the new proposals are set to the default values proposed by \citet{Hoffman14}.
A chain length of 1500 was simulated, with the first 500 iterations discarded as burn-in.
The sampler is initialized with the expectation value from the fit of the linear model with constant coefficients $\mu^{j}, \varsigma^{j}$ (Sect.~\ref{sec:ConstMod}) and assuming a value of one for all sills and a value of 60 km for all ranges of the GRF.
While the initialization is identical for all covariance parameters, the posterior estimates of the GRFs show distinguishable spatial dependency structures, depending on the parameter and model in question. 
Therefore, it can be concluded that it is fair to use the same priors and initialization for all covariance structures, as differences can successfully be inferred from the data.
Furthermore, all parameters converge to stable distributions within the first 500 iterations, indicating that the Markov chains have reached convergence and that the resulting posterior samples can be assumed as representative of their limit distributions.

All spatio-temporal predictors are standardized before training. 
They enter the parameter estimation process as deviations from the spatio-temporal mean, expressed in units of the standard deviation of the entire training data set. 
This normalization is performed with respect to the spatio-temporal mean rather than the local station means to preserve the spatial information contained in the local climatologies of the predictor variables.

\subsection{Spatio-temporal prediction} \label{sec:Prediction}

\subsubsection{Predictive distribution of gusts}
\label{sec:GustSampling}
In the Bayesian framework, probabilistic predictions are generated via a posterior predictive distribution. This is the conditional distribution of new observations, given the training observations.
The posterior predictive distribution can be derived from the posterior distributions of the parameters through integration over the uncertainties.
In practice for a BHM, this integration is achieved by a multi-stage sampling strategy \citep[cf.][Chapter 5]{Gelman14}.

In our gust model, the posterior distributions of the parameters are represented by the MCMC samples.
The uncertainties associated with the parameter estimates are integrated by iteratively drawing from the posterior samples, using a target sample size of $N=10.000$.
For each $k=1,\ldots,N$, one regression coefficient vector $\vec{\beta}_{i}$ is drawn from the MCMC samples, for location $i$.
Subsequently, the Gumbel parameters $\mu_{it},\varsigma_{it}$ are calculated via the regressions specified in Eq.~(\ref{eq:ConstMod}) and the current local observations of the predictor variables $\vec{x}^{\mu/\varsigma}_{it}$.
Ultimately, a single wind gust value is sampled from the resulting Gumbel distribution.
This procedure is repeated $N$ times to generate a predictive sample of wind gusts, conditional on location and predictor variables.
In consequence, the predictive distribution, as represented by the predictive sample, describes the whole uncertainty inherent to the model predictions.
From the predictive samples, predictive quantiles and threshold excess probabilities are calculated via the empirical distribution function, a method which is consistent with our verification framework outlined in Sect. \ref{sec:Verification} \citep{Krueger21}.
The quantiles are calculated using the median-unbiased method, as defined by \citet{Hyndman96}.
The threshold excesses are calculated as the fraction of draws from the posterior distribution that exceed the threshold. As our sample size is large, the effect of the ensemble size on the estimation of the threshold excess probabilities is negligible \citep{Ferro14}.

\subsubsection{Spatial interpolation of model parameters}
\label{sec:Kriging}
In order to make predictions at $p$ unobserved locations $\vec{s}_l, l=1,\ldots,p$, the spatial fields for the regression coefficients $\mu^j(\vec{r})$ and $\varsigma^j(\vec{r})$ need to be simulated at the respective prediction location, while conditioning on the parameter estimates at the training locations $\vec{r}_i, i=1,\ldots,q$.
The conditional simulation yields a multivariate Gaussian distribution for the prediction locations.

Following \citet{Stein99}, we can write the joint distribution of $\vec{\vartheta}_\mathrm{r}$ and the predictions $\vec{\vartheta}_\mathrm{s} = \mu^{j}/\varsigma^{j}(\vec{s}_1, \ldots,\vec{s}_p)$ as
\begin{equation}
    \binom{\vec{\vartheta}_\mathrm{s}} {\vec{\vartheta}_\mathrm{r}} \sim \mathcal{N} \left[ \binom{\vec{\alpha}_\mathrm{\vartheta}}{\vec{\alpha}_\mathrm{\vartheta}}, 
    \begin{pmatrix}
        \boldsymbol{\mathrm{\Sigma}}_\mathrm{s} & \boldsymbol{\mathrm{\Sigma}}_\mathrm{sr}\\
        \boldsymbol{\mathrm{\Sigma}}_\mathrm{rs} & \boldsymbol{\mathrm{\Sigma}}_\mathrm{r}
    \end{pmatrix} \right].
\end{equation}
The covariance matrices $\boldsymbol{\mathrm{\Sigma}}_\mathrm{r}$, $\boldsymbol{\mathrm{\Sigma}}_\mathrm{s}$, and $\boldsymbol{\mathrm{\Sigma}}_\mathrm{rs}$ are calculated using the Bayesian parameter estimates $\sigma_{\mu^{j}/\varsigma^{j}}$ and $\rho_{\mu^{j}/\varsigma^{j}}$ corresponding to the regression coefficient in question and the Mat\'ern covariance function from Eq.~(\ref{eq:maternLoc}). 
$\boldsymbol{\mathrm{\Sigma}}_\mathrm{r}$ is the covariance matrix at the observed locations, $\boldsymbol{\mathrm{\Sigma}}_\mathrm{s}$ the covariance at the predicted locations and $\boldsymbol{\mathrm{\Sigma}}_\mathrm{rs}=\left(\boldsymbol{\mathrm{\Sigma}}_\mathrm{sr}\right)^T$ represents the cross-covariance matrix between the estimated locations and the predicted locations.
Representations of the predictive parameters are obtained by sampling from $\mathcal{N}(\vec{\vartheta}_\mathrm{s|r}, \boldsymbol{\mathrm{\Sigma}}_\mathrm{s|r})$, where
\begin{equation}
 \vec{\vartheta}_\mathrm{s|r}= \vec{\alpha}_\mathrm{\vartheta} + \boldsymbol{\mathrm{\Sigma}}_\mathrm{sr}\boldsymbol{\mathrm{\Sigma}}_\mathrm{r}^{-1} (\vec{\vartheta}_\mathrm{r}- \vec{\alpha}_{\mathrm{\vartheta},0})
\end{equation}
represents the estimate of the conditional mean and
\begin{equation}
 \boldsymbol{\mathrm{\Sigma}}_{\mathrm{s|r}}=\boldsymbol{\mathrm{\Sigma}}_\mathrm{s}-\boldsymbol{\mathrm{\Sigma}}_\mathrm{sr}\boldsymbol{\mathrm{\Sigma}}_\mathrm{r}^{-1}\boldsymbol{\mathrm{\Sigma}}_\mathrm{rs}
\end{equation}
the conditional covariance matrix.
This process is repeated for all 1000 elements of the MCMC samples to obtain a posterior sample of the spatial parameter fields at the prediction locations.
This posterior sample is then used for the multi-stage sampling strategy outlined in Sect. \ref{sec:GustSampling}.

\subsection{Verification}
\label{sec:Verification}
The resulting predictive  distributions are compared to observation values and evaluated by scoring rules to assess the models' predictive quality.
To this end, we apply cross-validation in space and time.
In time, the data is separated in two approximately equal-sized data sets, one for training and one for evaluation.
This separation ensures that all predictions used for model verification are out-of-sample with respect to the training period.
To account for possible autocorrelation and an annual cycle, the data set is partitioned by year, with odd-numbered years designated for training and even-numbered years for verification. 
To guarantee that the spatial prediction is out-of-sample, that is, that no data from the verifying location was used for model training, we employ leave-one-out cross-validation by always withholding the data for one station.
Subsequently, the $\mu^{j}/\varsigma^{j}$ are predicted in space for the withheld location and the wind gusts are predicted using the days from the evaluation years.
This way, time series of probabilistic predictions and observations are generated for all stations and dates.

The post-processing models are evaluated and compared by out-of-sample predictive performance.
Therefore, the verification is conducted via proper scoring rules \citep{Gneiting07}.
We use Brier score \citep{Brier50} for the verification of threshold exceedance probabilities and  the quantile score \citep{Koenker05, Koenker99, Friederichs07}, also referred to as general piecewise linear score \citep{Gneiting23}, for the verification of predictive quantiles.
Brier score (BS) is given by \citep{Wilks19}
\begin{equation}\label{eq:BrierScore}
\operatorname{S}_\mathrm{B}^u(F,y) = (p_u - \mathbb{I}\{y > u\})^2,
\end{equation}
and quantile score (QS) is given by \citet{Bentzien14} as
\begin{equation}\label{eq:QuantileScore}
 \operatorname{S}^{\tau}_\mathrm{Q}(F,y) = \rho_\tau(y-F^{-1}(\tau)),
\end{equation}
where $\rho_\tau(u) = \tau u$ if $u\geq 0$, and $\rho_\tau(u) = (\tau-1) u$ otherwise.
Both scores are negatively oriented, so that lower values imply better predictive performance.

The decomposition of the scoring rules into uncertainty, reliability and resolution components is also calculated \citep{Murphy73, Broecker09}.
We refer to the decomposition components as ``uncertainty'', ``miscalibration'' and ``resolution''. 
With respect to miscalibration (MCB), we follow \citet{Gneiting23}.
In Murphy's original decomposition \citep{Murphy73}, a lower reliability value implies a higher reliability of the model prediction. Hence, to avoid semantic ambiguity, we use the term ``miscalibration'' instead.
With the term ``resolution'' (RES), we adhere to the standard nomenclature within the calibration-refinement factorization of the joint distribution of forecasts and observations \citep{Wilks19}.
The uncertainty (UNC) only depends on the observations and is defined as the score of the climatology forecast and quantifies the data-immanent difficulty to forecast the variable in question.
UNC and MCB are negatively oriented, so that higher values correspond to lower skill. RES is positively oriented, so that higher values correspond to larger skill.
The total score $\operatorname{S}$ can be decomposed as 
\begin{equation}\label{eq:ScoreDecomp}
    \operatorname{S} = \operatorname{MCB}-\operatorname{RES}+\operatorname{UNC}.
\end{equation}

Estimating the decomposition from real-valued data presents challenges regarding the binning of the forecast-observation pairs \citep{Bentzien14, Atger04}.
In this work, we follow the approach by \citet{Dimitriadis21}, who develop a stable estimation of the decomposition terms of BS that produces consistent, optimally binned, reproducible and PAV (pool adjacent violators)-based (CORP) results.
Similarly, the CORP approach for general metrics, including QS, was derived by \citet{Jordan22}.
This procedure performs an isotonic regression on the forecast and observation values using the pool-adjacent-violators algorithm  \citep[PAVA,][]{Ayer55, Miles59, Best90, deLeeuw09}, thereby conditionally calibrating the forecast values on the observations. 
For BS, the threshold exceedance forecasts are calibrated to the mean of the corresponding binary-converted observations. For QS, the forecast values are calibrated to the respective quantile functional of the corresponding gust observations.
Miscalibration is calculated from the original score $\operatorname{S}$ and the recalibrated score $\operatorname{S}_\mathrm{rc}$ as
\begin{equation}
    \operatorname{MCB} \equiv \operatorname{REL} =: \operatorname{S} - \operatorname{S}_\mathrm{rc},
\end{equation}
and the discrimination/resolution part is obtained from the recalibrated score and the marginal score as
\begin{equation}
    \operatorname{RES} \equiv \operatorname{DSC} =: \operatorname{S}_\mathrm{rc} - \operatorname{S}_\mathrm{mg} \equiv \operatorname{S}_\mathrm{rc} - \operatorname{UNC},
\end{equation}
where the marginal score $\operatorname{S}_\mathrm{mg}$ refers to the score obtained from the climatology forecast \citep{Gneiting23}. The marginal score is identical to the uncertainty.
In this work, we follow the CORP-based QS decomposition from \citet{Gneiting23} and the BS decomposition from \citet{Dimitriadis21}, which is implemented in R \citep{R} in the package \verb+reliabilitydiag+.

The skill of a model can be assessed through derived skill scores \citep{Wilks19}.
Skill scores are obtained from a scoring rule $…\operatorname{S}$ via 
\begin{equation}\label{eq:SkillScore}
    \operatorname{SS} = \frac{\operatorname{S}-\operatorname{S}_\mathrm{ref}}{\operatorname{S}_\mathrm{perf}-\operatorname{S}_\mathrm{ref}} = 1 - \frac{\operatorname{S}}{\operatorname{S}_\mathrm{ref}},
\end{equation}
where $\operatorname{S}_\mathrm{perf}$ refers to the perfect score. 
The simplification on the right-hand side of Eq.~(\ref{eq:SkillScore}) is possible, as both BS and QS have a perfect score of $\operatorname{S}_\mathrm{perf}=0$. 
If the climatology is used as reference, the skill score can be obtained from the decomposition components as
\begin{equation}
\label{eq:SkillFromDecomp}
    \operatorname{SS} = \frac{\operatorname{RES}-\operatorname{MCB}}{\operatorname{UNC}}.
\end{equation}
Skill scores are positively oriented, indicating higher skill with higher values. 
The perfect forecast has a value of 1. Positive values indicate superior skill to the reference model, whereas negative values indicate lower skill. 
If we refer to skill against climatology in this work, we refer to a locally calculated climatology of the gust observations at each station.

\subsubsection{Verification of the posterior distribution}

Additionally, we assess the quality of the parameter estimation using the continuous ranked probability score \citep[CRPS, ][]{Hersbach00}. CRPS evaluates the whole probability distribution and can be regarded as either the integral of BS over all real-numbered thresholds \citep{Hersbach00} or the integral of QS over all possible quantiles \citep{Gneiting11}. We use CRPS in the form given by \cite{Gneiting07} as
\begin{equation}
    \label{eq:CRPS}
    \operatorname{S}_\mathrm{CRP} = \mathbb{E}_\mathrm{F}|Y-o| - \frac{1}{2}\mathbb{E}_\mathrm{F}|Y-Y^\prime|,
\end{equation}
for forecast $Y$ and observations $o$, with $Y^\prime$ being an independent random copy of $Y$. The expectation values $\mathbb{E}_\mathrm{F}$ can be approximated numerically or calculated in closed form under distributional assumptions. As the marginal posterior distributions of our gust model are sufficiently close to Gaussian, we approximate CRPS using the closed form CRPS \citep{Gneiting05}
\begin{equation}
    \label{eq:CRPSGauss}
    \operatorname{S}_\mathrm{CRP}(\mu,\sigma^2,o)=\sigma\left\{\frac{o-\mu}{\sigma}\left[2\Phi\left(\frac{o-\mu}{\sigma}\right)-1\right]+2\phi\left(\frac{o-\mu}{\sigma}\right)-\frac{1}{\sqrt\pi}\right\}
\end{equation}
for mean $\mu$ and variance $\sigma^2$ of a Gaussian distribution. $\Phi$ and $\phi$ denote the cumulative distribution function and the probability density function of the standard normal distribution.\\
For a joint evaluation of the posterior distribution for the regression coefficients, we apply the energy score \citep[ES,][]{Gneiting07}, which is the multivariate extension of the CRPS. ES is given as
\begin{equation}
    \label{eq:EnergyScore}
    \operatorname{S}_\mathrm{E} = \mathbb{E}_\mathrm{F}||\vec{Y} -\vec{o}|| - \frac{1}{2}\mathbb{E}_\mathrm{F}||\vec{Y} - \vec{Y}^\prime||,
\end{equation}
where $||\cdot||$ denotes an euclidean distance. We scale the regression coefficients by the inverse of their spatial variance obtained from maximum likelihood fits on the data from each station to ensure that they contribute homogeneously to the distance. In the case of ES, we approximate the expectation values $\mathbb{E}_\mathrm{F}$ numerically by averaging over all representations from the MCMC samples.

\section{Results}\label{sec:Results}
\subsection{Spatially constant and local models}
\label{sec:ConstMod}
The benefit of \mbox{SpatBHM} for wind gust post-processing is assessed against a spatially constant baseline model (ConstMod), where all regression coefficients $\vec{\mu}^{j}$ and $\vec{\varsigma}^{j}$ are spatially constant. 
Although \mbox{ConstMod} is technically not a Bayesian hierarchical model, we treat it as such and use the same Bayesian inference methods as for \mbox{SpatBHM}.
To ensure consistency in the model formulation, the priors of the corresponding GRF mean values are assigned to the constant regression coefficients of \mbox{ConstMod}.
Provided that the covariates are available at the prediction location, \mbox{ConstMod} also allows the post-processing of wind gusts at unobserved locations, since it is assumed that the regression coefficients do not vary in space in this case.

In search of a strong baseline, we conducted a selection of predictors and predictand for the linear model, based on predictive skill (Appendix \ref{A:PredictorSelection}). 
Models using $\mathrm{FX}-V_\mathrm{m}$ as predictand showed higher skill as models predicting $\mathrm{FX}$ directly.
By testing various combinations of predictors, we find the best linear model for the location $\mu_{it}$ and the scale $\sigma_{it}$ parameter as
\begin{eqnarray}\label{eq:ConstMod}
 \mu_{it} &=&\mu^{0} + \mu^{1}   V_{\mathrm{max},it}  + \mu^{2} V_{\mathrm{m},it} + \mu^{z} \Delta z_i\notag \\ 
 \log( \varsigma_{it} ) & = &\varsigma^{0} + \varsigma^{1}   V_{\mathrm{max},it} + \varsigma^{2} V_{\mathrm{m},it},    
\end{eqnarray}
for locations $i = 1, \ldots,q$ and times $t = 1, \ldots, n$, and predicting $\mathrm{FX}-V_\mathrm{m}$. 
$V_\mathrm{max}$ represents the maximum wind speed in COSMO-REA6, $V_\mathrm{m}$ represents the mean horizontal wind from COSMO-REA6 and $\Delta z_i$ is a spatial covariate for the influence of the station altitude, given by $\Delta z_i =  z_{\mathrm{synop},i}-z_{\mathrm{COSMO},i}$.
For the training of \mbox{ConstMod}, we simulated Markov chains of length $N=1250$, using only 250 iterations as burn-in because of the lower complexity of the model.

The results are further contextualized by comparing them to the results from a local model (\mbox{LocMod}), where individual regression coefficients are estimated at each station. While in the cases of \mbox{ConstMod} and \mbox{SpatBHM}, the local data are excluded for each station in cross-validation, \mbox{LocMod} is specifically trained on the local data only. Consequently, we expect \mbox{LocMod} to represent the maximum attainable skill from the data with the type of linear model utilized.

\begin{figure}
    \centering
    \includegraphics[width=\textwidth]{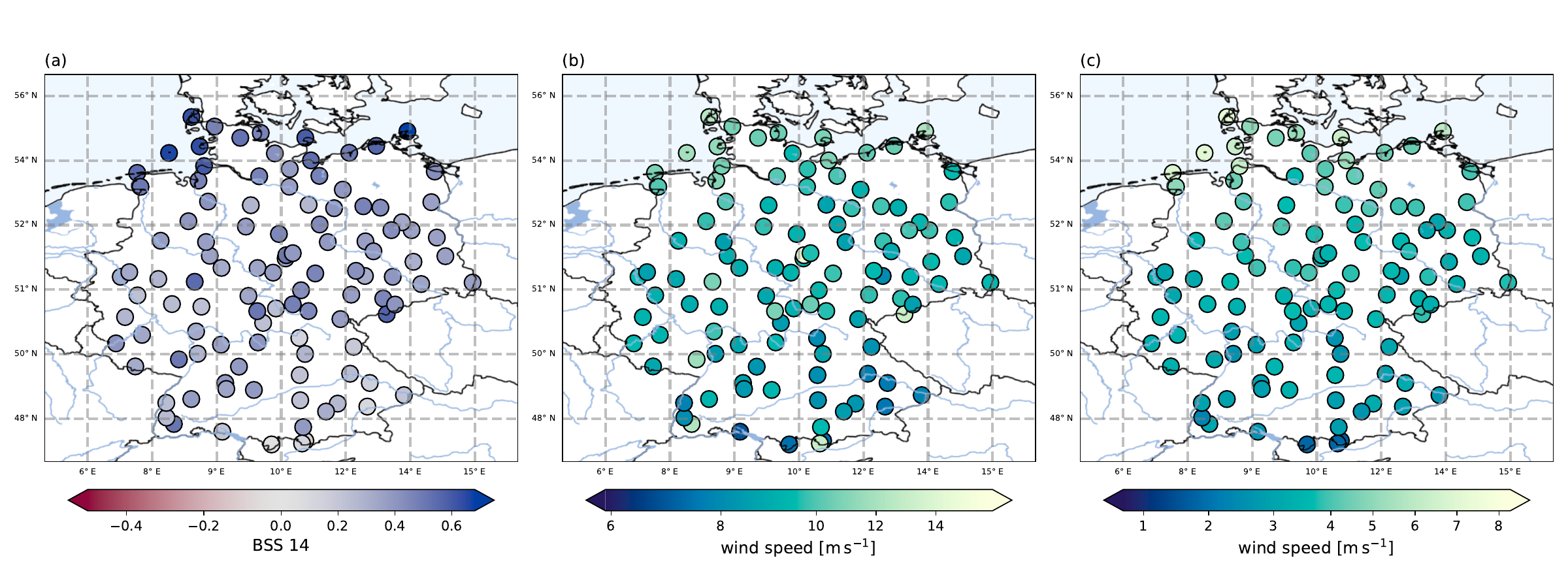}
    \caption{(a) Spatial distribution of the skill of LocMod against climatology for exceedances of the 14 \unit{m~s^{-1}} threshold (BSS 14). Blue colors indicate positive skill, red colors indicate negative skill. Each dot is one SYNOP station used for training. (b) Mean $\mathrm{FX}$ value used for training. (c) Mean $V_\mathrm{m}$ value used for training.}
    \label{fig:SkillLocMod}
\end{figure}

\begin{figure}
    \centering
    \includegraphics[width=\linewidth]{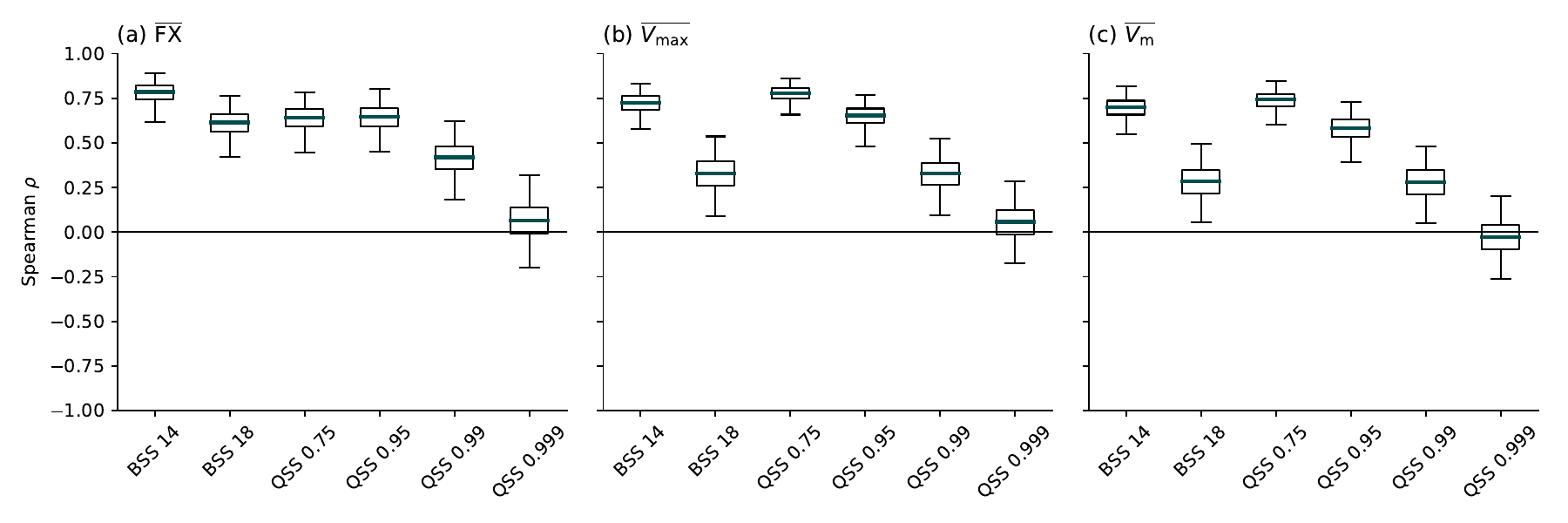}
    \caption{Spearman rank correlation coefficient between various variables and LocMod skill against climatology, shown for exceedance probabilities of the 14 \unit{m~s^{-1}} (BSS 14) and 18 \unit{m~s^{-1}} (BSS 18) threshold, and the 0.75 to 0.999 quantiles (QSS 0.75 to QSS 0.999). Correlations are shown (a) for the gust observations $\mathrm{FX}$, (b) the maximum wind speed $V_\mathrm{max}$ and (c) the mean wind $V_\mathrm{m}$. Boxes represent the interquartile range and whiskers extend to the 0.01 and 0.99-quantiles. Uncertainty estimates are based on 5000 bootstrap iterations. In each iteration, both the dates for the calculation of the mean wind and the stations to calculate the correlation are resampled.}
    \label{fig:CorrWindSkill}
\end{figure}

As visible from the map in Fig.~\ref{fig:SkillLocMod}a, even \mbox{LocMod} demonstrates negative skill against the climatology at some locations. The stations with negative skill are mostly situated in mountainous terrain, except for QSS 0.999, where a more general decrease in skill is observed (not shown).
This suggests that the COSMO-REA6 predictor data lack explanatory power for the wind gust observations at these stations, pointing to deficiencies in the representation of topography in COSMO-CLM.
Additionally, there is a coastward gradient with lower possible skill in southern Germany and higher skill near the coast. 

Figure~\ref{fig:CorrWindSkill} shows that the maximum attainable skill depends on the local mean wind characteristics. LocMod skill against climatology is strongly correlated with the mean wind at each station over all investigated threshold exceedance probabilities and quantiles, except for the most extreme quantile. 
This mean wind dependency of the attainable skill leads to a more successful post-processing at windy locations, which explains why the highest level of attainable skill is found near the coastlines and in mountainous areas.
Particularly, the station with lowest attainable skill is Garmisch-Partenkirchen, which is also the station recording the lowest average gust speed among the 109 locations.

\begin{figure}
    \centering
    \includegraphics[width=\textwidth]{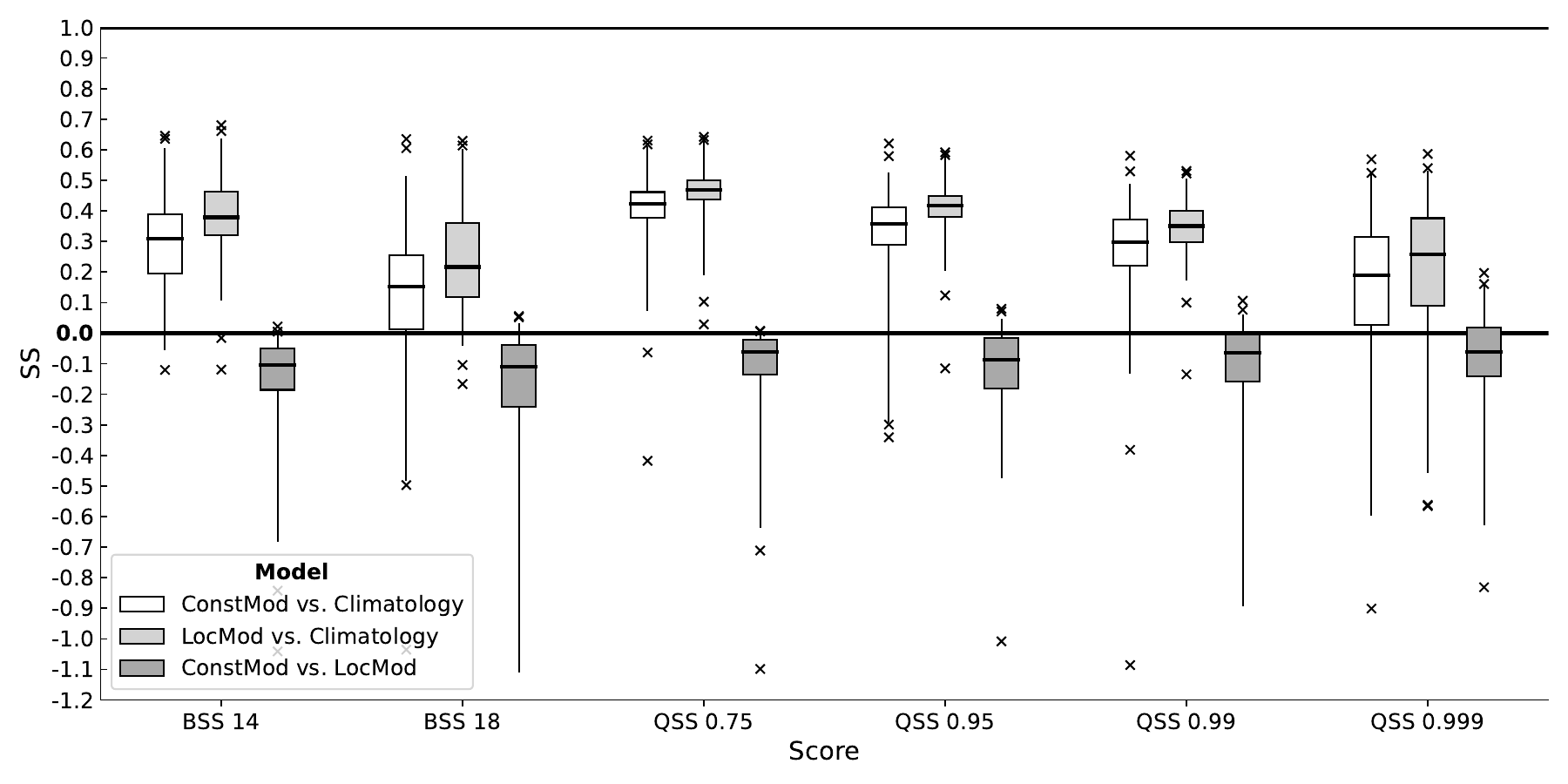}
    \caption{Cross-validated skill scores of \mbox{ConstMod} and \mbox{LocMod} against climatology, and \mbox{ConstMod} against \mbox{LocMod} for exceedance probabilities of the 14 \unit{m~s^{-1}} (BSS 14) and 18 \unit{m~s^{-1}} (BSS 18) thresholds, and the 0.75 to 0.999 quantiles (QSS 0.75 to QSS 0.999). Each box plot contains skill scores at 109 locations. Boxes represent the interquartile range and whiskers extend to the $0.01$ and $0.99$-quantiles. Bold lines mark the median skill scores. Median score values are given in Table~\ref{tab:ConstmodOverview}.}
    \label{fig:SkillConstClimLocal}
\end{figure}

Figure~\ref{fig:SkillConstClimLocal} shows the cross-validated skill scores for \mbox{ConstMod} compared to the local climatology and LocMod for the 14 and 18 \unit{m~s^{-1}} threshold \citep[cf.][]{DWDwarnings} and the $0.75$, $0.95$, $0.99$ and $0.999$-quantiles obtained at all investigated locations. 
\mbox{ConstMod} demonstrates a high level of skill against the local climatology. 
The skill is high over all investigated distributional features with average skill score values between 10 \unit{\%} and 45 \unit{\%} and approaches the skill of \mbox{LocMod}.
The highest median skill is found for the 0.75-quantile at values greater than 40 \unit{\%}. 
The 0.75-quantile also has the least spread in skill, implying that the prediction skill for this quantile is spatially homogeneous.
For higher thresholds and outer quantiles, we observe less skill and also locations with negative skill against climatology. However, the number of instances with negative skill remains similar to the instances with negative skill found from \mbox{LocMod}.
Moreover, the 18 \unit{m~s^{-1}} threshold is rarely exceeded within the data set, resulting in a large dispersion of BSS values. 
Likewise, the 0.999-quantile is extreme and therefore suffers from the generally low predictability of extreme events.
The comparison to \mbox{LocMod} shows that \mbox{ConstMod} already closely approaches the maximum attainable skill, and therefore offers an excellent baseline model that is hard to beat.

\subsection{Effect of the elevation offset}
\label{Sec:altitude}

\begin{figure}
    \centering
    \includegraphics[width=\textwidth]{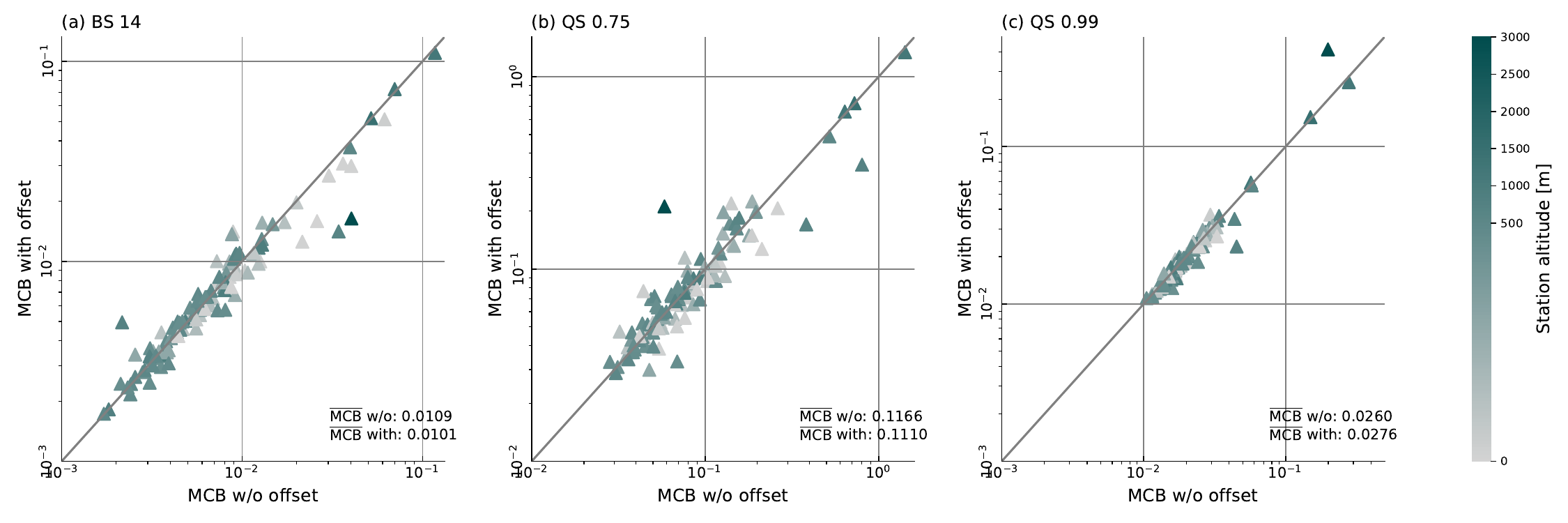}
    \caption{Miscalibration of \mbox{SpatBHM} with and without the elevation offset in the covariance function. Each data point represents the miscalibration at one station. The coloring represents station altitude of the station in question. Miscalibration values are shown for (a) BS 14, (b) QS 0.75, and (c) QS 0.99. The gray line represents the identity function. Note that the axes are scaled logarithmically, so that smaller values are overrepresented.} 
    \label{fig:EffectAltitude}
\end{figure}

We evaluate the benefit of including the elevation offset from Eq.~(\ref{eq:AltitudeScaling}) in the distance metric for the covariance calculation in a simple version of \mbox{SpatBHM}. In this simple version only one GRF is assigned to $\mu^0$, while the rest of the regression coefficients are kept spatially constant.
The model is trained once using Eq. (\ref{eq:GreatCircleDist}) as distance metric and once using Eq. (\ref{eq:AltitudeEuclidean}).
As a result of the elevation offset, the estimated values for the range parameter $\rho_{\mu^{j}}$ for \mbox{SpatBHM} become larger and show higher stability during cross-validation.
The larger range parameter values lead to a smoother spatial dependency structure, as the elevation offset effectively mitigates the impact of the mountain stations on their direct neighbors. This enhances the discernibility of the large-scale spatial dependency structure.

The added value in terms of miscalibation for \mbox{SpatBHM} is displayed in Fig.~\ref{fig:EffectAltitude}. 
For BS 14 (Figure~\ref{fig:EffectAltitude}a), the elevation offset has a positive impact on the calibration of SpatBHM, which is improved particularly for stations with large MCB. 
For QS 0.75 (Figure~\ref{fig:EffectAltitude}b),  we see a similar effect, although it is not as pronounced as for BS 14. There are cases where the introduction of the elevation offset has a negative impact, but these stations generally show a good calibration, regardless of the elevation offset.
For the more extreme 0.99 quantile (Figure~\ref{fig:EffectAltitude}c), mean MCB is higher than without the offset, but this is mainly caused by the deterioration at one single station, i.e. Zugspitze. 
Since this station repeatedly appears as an outlier in our investigation, one possible explanation is that the Zugspitze station has a significantly different gust climatology compared to other locations and might be poorly represented by its corresponding grid cell in COSMO-REA6.

In summary, the elevation offset enables the mountain stations to be retained in the training data set and mostly improves the calibration of the model without unduly influencing the model skill at already well represented locations.

\subsection{Predictive performance of SpatBHM} \label{sec:SkillSpatBHM}

We conducted a non-exhaustive model selection in order to find a good combination of spatial fields for the parameters (Appendix \ref{A:SpatialFields}).
For each GRF, all local representations at the training locations have to be estimated, leading to a significant increase in the number of parameters. 
Thereby, the computational burden for the parameter estimation increases substantially with more GRFs, without necessarily improving the predictive skill.
We decided for a compromise using a model with only a limited number of spatial parameters that shows high skill, while keeping the computations for the parameter estimation practicable. 
The \mbox{SpatBHM} version that was found best has two spatial regression coefficients: the intercept for the location parameter $\mu^{0}$ and the regression coefficient for $V_\mathrm{m}$ for the location parameter $\mu^{2}$. 
Please note that $\mu^{3}$ which controls a purely spatial covariate, is always assumed as spatially constant.
There is limited evidence that a spatial field for the scale parameter improves the estimation of the threshold exceeedance probabilities. 
However, it also leads to a significant loss in QS, which is the reason why it was discarded. The fitted values of the model parameters are shown in Table~\ref{tab:SpatbhmParvalues}.

\begin{figure}
    \centering
    \includegraphics[width=\linewidth]{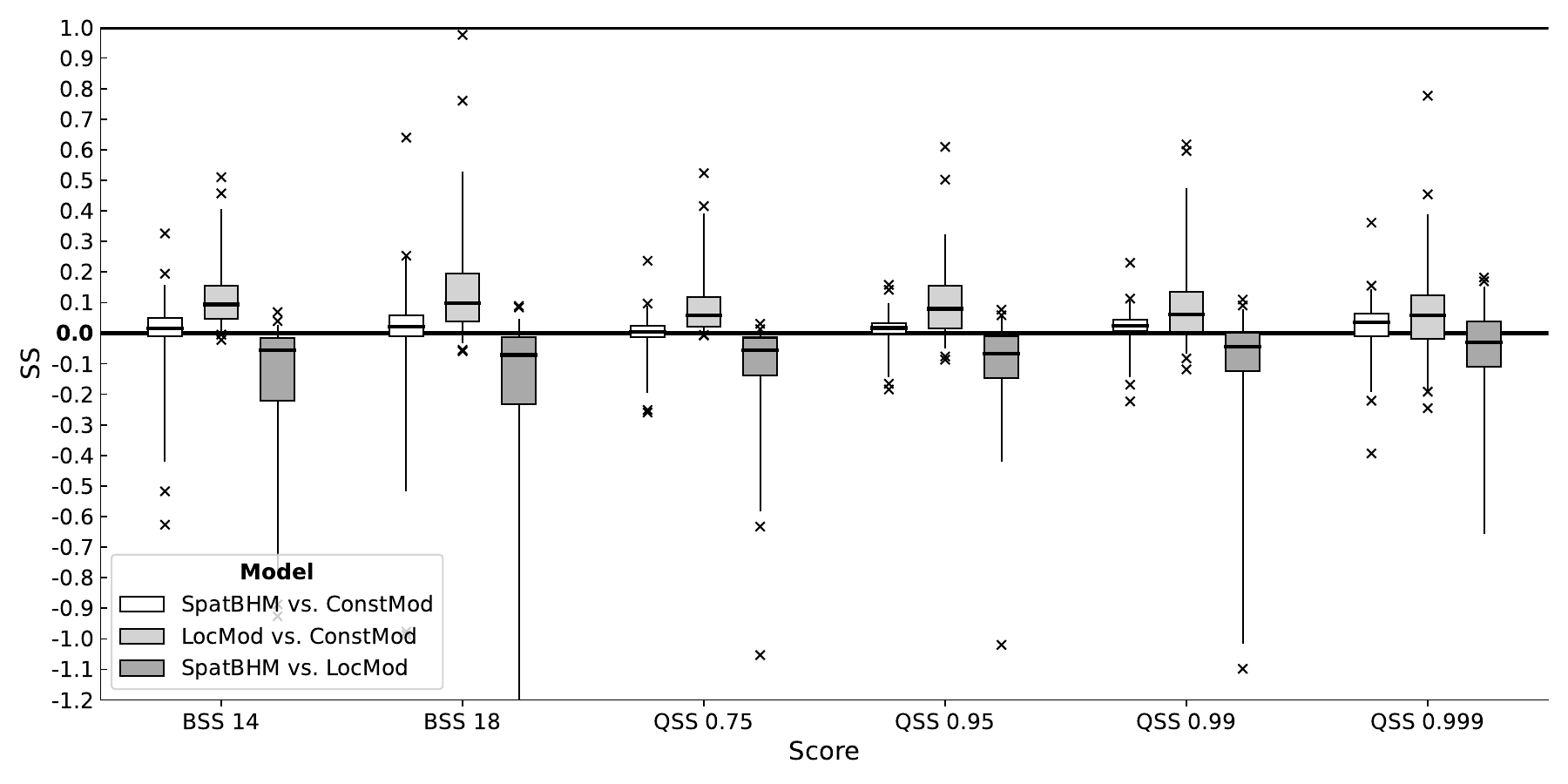}
    \caption{As Fig. \ref{fig:SkillConstClimLocal} but for cross-validated skill scores against of SpatBHM and LocMod against ConstMod and SpatBHM against LocMod.} 
    \label{fig:SkillSpatConstLocal}
\end{figure}

Figure \ref{fig:SkillSpatConstLocal} shows the skill score values of SpatBHM against ConstMod for the 14 and 18 \unit{m~s^{-1}} threshold and the $0.75$, $0.95$, $0.99$ and $0.999$-quantiles obtained at all investigated locations.
The BSS values of \mbox{SpatBHM} compared to \mbox{ConstMod} for the threshold exceedance probabilities  are close to 0.
However, the median and the larger part of the interquartile range are found above 0, pointing to a slight improvement in prediction quality compared to \mbox{ConstMod}.
The median BSS is found around $2$  \unit{\%}. 
For the prediction quantiles, there is a clear improvement in skill, especially for the higher quantiles. 
The median skill score values range from 2--5  \unit{\%}, depending on the exact quantile. 
For the $0.95$ and $0.99$-quantiles, almost the whole interquartile range is positive and for the $0.999$-quantile, \mbox{SpatBHM} skill nearly achieves the same skill as \mbox{LocMod}.
At least for the representation of the prediction quantiles, the relative skill of \mbox{SpatBHM} compared to \mbox{ConstMod} can be regarded as a proof of concept of the spatial hierarchical approach.

However, there are some instances, where \mbox{SpatBHM} performs poorly compared to \mbox{ConstMod}. These are mostly stations, where the local wind gust characteristics change rapidly over space. In some cases, the spatial mean of the GRF is more representative for the local gust model than the coefficients at the surrounding stations. Therefore, the spatial interpolation procedure introduces a bias and the skill of \mbox{SpatBHM} is reduced compared to \mbox{ConstMod}. Nevertheless, the skill against the climatology forecast remains high at the majority of these locations.

Given the spatial model's notable predictive skill for higher quantiles, we proceed to evaluate its performance in stronger wind conditions.
Absolute extreme events are evaluated in terms of the exceedance probability of high thresholds.
Due to the scarcity of observations for high threshold exceedances, it is not reasonable to calculate BS for each station individually. Instead, we compute the average BS over all stations.
Thereby, we increase the number of observations and obtain interpretable values for BSS, which are shown in Fig. \ref{fig:BssStrat}.
As higher thresholds, the next three warning levels by the DWD, i.e. 25, 29 and 33 \unit{m~s^{-1}} \citep{DWDwarnings}, are selected.

\begin{figure}
    \centering
    \includegraphics[width=\textwidth]{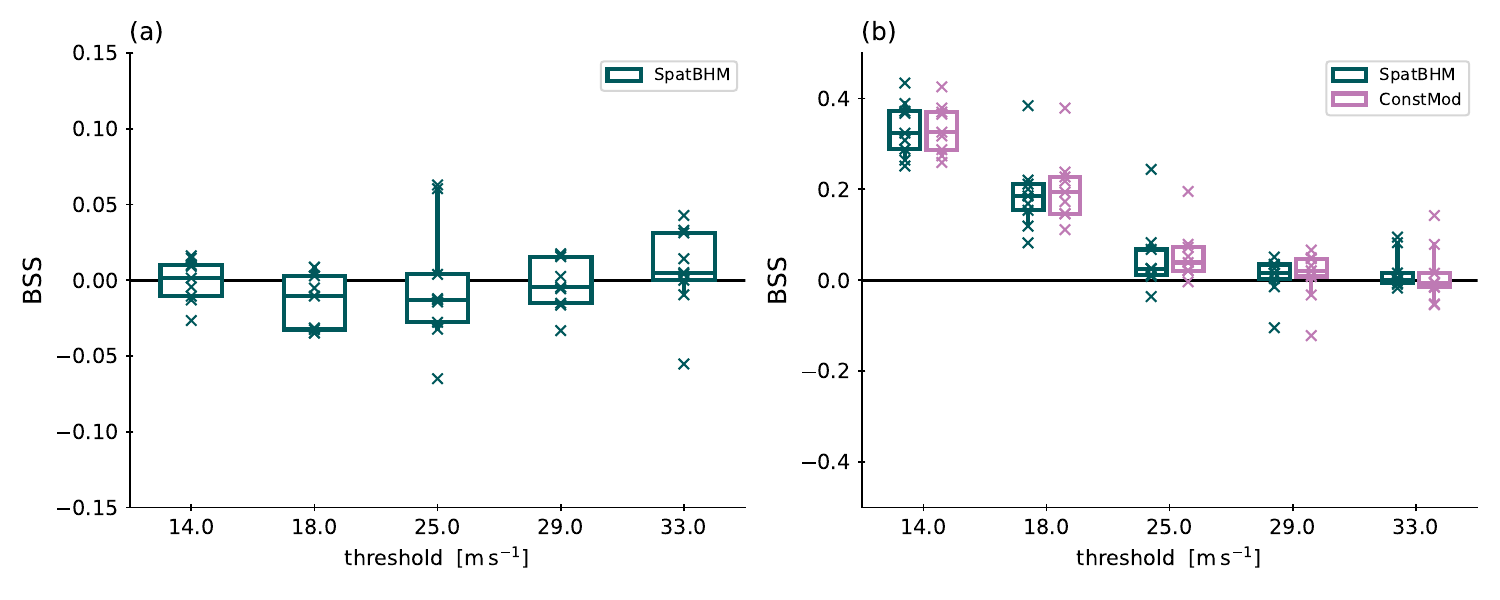}
    \caption{(a) BSS of \mbox{SpatBHM} against \mbox{ConstMod}, aggregated over all stations. Each boxplot comprises 9 values, one for each year included in the evaluation data set. The skill score values are calculated from the mean score at all stations throughout each year. (b) As in (a) but for \mbox{SpatBHM} and \mbox{ConstMod} against climatology.}
    \label{fig:BssStrat}
\end{figure}

\mbox{SpatBHM} performs very similar to \mbox{ConstMod} on the higher thresholds. Only for the highest threshold (33 \unit{m~s^{-1}}), it has visibly higher skill.
The skill is lowest for the $u=18$ \unit{m~s^{-1}} and $u=25$ \unit{m~s^{-1}} thresholds.
The skill against the climatology in Fig. \ref{fig:BssStrat}b provides context on the skill for the threshold exceedance probabilities: 
Owing to the extreme nature of the events, neither \mbox{SpatBHM} nor \mbox{ConstMod} show any significant skill against the climatology for the highest threshold, and the skill for all thresholds above 25 \unit{m~s^{-1}} is generally low.
In summary, the spatial approach does not significantly improve the prediction of threshold excess probabilities from the spatially constant approach, but neither does it perform significantly worse.

\begin{figure}
    \centering
    \includegraphics[width=\textwidth]{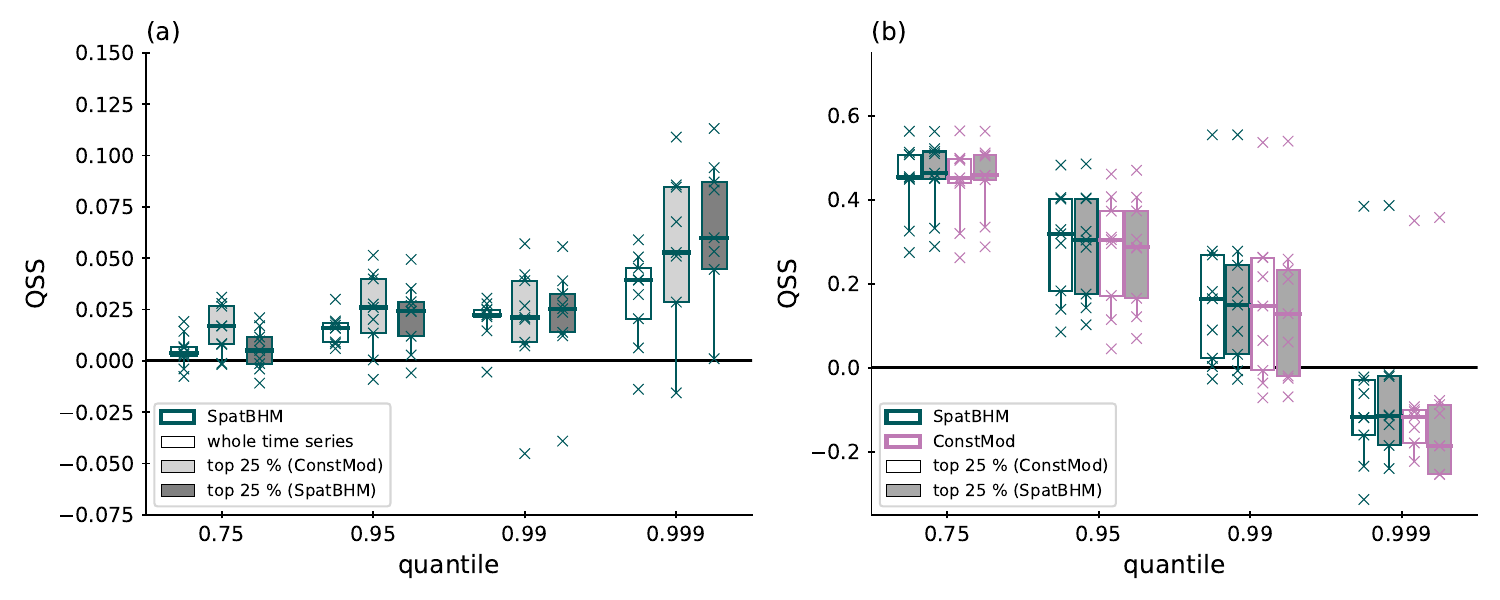}
    \caption{(a) QSS for \mbox{SpatBHM} against \mbox{ConstMod}. Each boxplot comprises one value for each of the 9 years included in the evaluation data set. The skill score values are aggregated over all 109 investigated locations. The white shaded boxes represent the average annual skill evaluated on the whole time series. The light and dark gray shaded boxes represent the skill, when only using the top 25  \unit{\%}, stratified on the highest respective quantile forecasts selected from \mbox{ConstMod} and \mbox{SpatBHM}. (b) Same for \mbox{SpatBHM} against climatology in comparison to \mbox{ConstMod}. The time series for verification are stratified based on the predictions from ConstMod (white) and SpatBHM (gray).}
    \label{fig:QssStrat}
\end{figure}

In order to quantify SpatBHM's performance for quantile predictions in higher wind conditions, we stratify the data and calculate QS for the largest 25 \unit{\%} of the forecast values.
The stratification is based on the forecast values, rather than the observations, in order to avoid the forecaster's dilemma \citep{Lerch17}.
The top 25 \unit{\%}-wind gust events are selected from the whole data set. The majority of these events is observed at coastal locations and mountain tops.
As the stratification significantly reduces the data at some stations, the score values are averaged over all locations.
We stratify once based on the predictions from \mbox{SpatBHM} and once based on the predictions from \mbox{ConstMod}.
For the estimation of uncertainties, score values are calculated for each year of the verification data set, resulting in a sample of 9 independent score estimations.
The sample size is insufficient to yield statistically robust results, but it provides an indication of the potential outcome. The resulting skill score values for \mbox{SpatBHM} and \mbox{ConstMod} are depicted in Fig.~\ref{fig:QssStrat}.

In contrast to the threshold exceedance probabilities, the prediction quantiles in Fig.~\ref{fig:QssStrat}a almost consistently exhibit higher skill for \mbox{SpatBHM} than for \mbox{ConstMod}, across all evaluated quantiles.
The spatial model exhibits considerably higher quantile skill for the top 25 \unit{\%} gust events than for the complete data set.
This skill improvement is found for both types of stratification, so it does not depend on the model used for the stratification, although there are minor differences. 
Similar to the separate evaluation for all locations, the spatially aggregated skill compared to \mbox{ConstMod} increases with higher quantile. 
Moreover, \mbox{SpatBHM} consistently outperforms \mbox{ConstMod}, independent of which of the two models was used to select the top 25 \unit{\%} of predictions.
Therefore, the spatial modeling approach can be considered to particularly improve the prediction of rare or extreme events in terms of prediction quantiles.
However, in this context it should be noted, that the overall skill of both models compared to the climatology decreases with higher quantile (Fig. \ref{fig:QssStrat}b). 

\begin{figure}
    \centering
    \includegraphics[width=\linewidth]{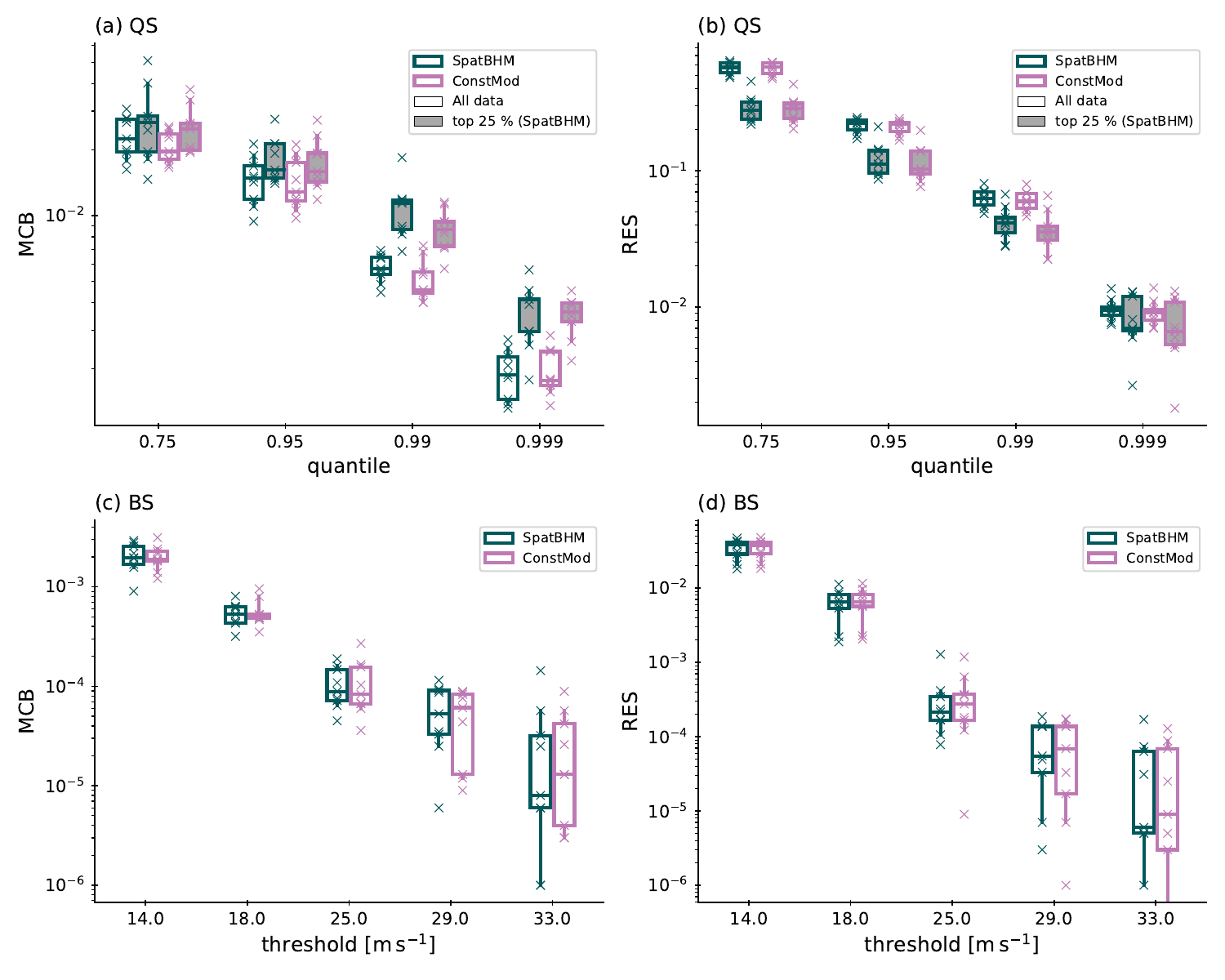}
    \caption{(a) QS miscalibration and (b) QS resolution for \mbox{SpatBHM} and \mbox{ConstMod} in comparison for evaluation on the complete data set and only the top 25 \unit{\%} of data. Results for stratification are based on \mbox{SpatBHM}. Specifications of box plots as in Fig.~\ref{fig:QssStrat}. (c) BS miscalibration and (d) BS resolution for \mbox{SpatBHM} (green) and \mbox{ConstMod} (pink). The score decomposition is calculated over the full time series without stratification.}
    \label{fig:RelResStrat}
\end{figure}

For a better understanding of the origin for the improved skill, Fig.~\ref{fig:RelResStrat} shows the score decomposition from Eq.~(\ref{eq:ScoreDecomp}) for BS and QS.
The inter-model differences of MCB and RES for the prediction of threshold exceedance probabilities (cf. Fig. \ref{fig:RelResStrat}c and d) are small.
MCB is better in \mbox{SpatBHM} for the higher thresholds, whereas RES is better in \mbox{ConstMod} for the higher thresholds.
However, the inter-model differences are more distinctive for the prediction quantiles in Fig. \ref{fig:RelResStrat}a and b. MCB is higher for \mbox{SpatBHM} and lower for \mbox{ConstMod}, pointing to poorer calibration in the spatial approach. 
The $0.999$-quantile is an exception in that MCB is similar for both models and slightly better for \mbox{SpatBHM} for both the complete and reduced data sets.
A possible reason for the inferior calibration of \mbox{SpatBHM} compared to \mbox{ConstMod} is the spatial interpolation procedure.
For locations with wind gust characteristics similar to the spatial mean, \mbox{ConstMod} already has high skill and the spatial approach would not be required.
If a station exhibits distinct wind gust characteristics from its surrounding stations, the kriging process results in a value too close to nearby observations and introduces a bias.
This bias can only partially be removed by the elevation offset in Eq.~(\ref{eq:AltitudeScaling}).
Consequently, \mbox{SpatBHM} only exhibits better calibration at locations, where the wind gust characteristics diverge from the spatial mean, yet align with the surrounding stations.
Therefore, the calibration of \mbox{SpatBHM} is reduced compared to \mbox{ConstMod}.
RES, shown in Fig. \ref{fig:RelResStrat}b and d, is very similar for both models. RES is slightly better for \mbox{SpatBHM}, most visibly for the extreme quantiles. As RES values are an order of magnitude larger than MCB, the resolution part of the score is what defines the local skill. Therefore we conclude that \mbox{SpatBHM} outperforms \mbox{ConstBHM} not due to better calibration but rather due to a better resolution. 

In summary, our results can be regarded as a proof of concept, that the spatial hierarchical approach can contribute to an improved wind gust post-processing from reanalysis data, as it provides on average more spatially homogeneous estimates of the observed wind gusts.
Our stratified analysis showed that the increase in skill is especially noteworthy for higher prediction quantiles and in higher wind conditions, which points to a better representation of the variability of the wind gusts.
Nevertheless, these results do not generalize to locations with low average windiness with low \mbox{LocMod} skill against climatology.
Consequently, the spatial probabilistic post-processing provides high skill for coastal stations and some mountain tops, and less skill for valley locations.
The high skill at coastal stations, however, suggests that the spatial model successfully detects and corrects systematic regional gust behavior deviations.\\

\subsection{Quality of the spatial parameter prediction}

To assess the added value of \mbox{SpatBHM} over \mbox{ConstMod} with regard to the spatial prediction of the model parameters, we calculate the CRPS for the marginal posterior predictive distributions of the regression coefficients $\mu^{i}$ and $\varsigma^{i}$.  As observations, the parameter estimations from the \mbox{LocMod}-fits are used (Sect.~\ref{sec:ConstMod}).
In the case of \mbox{ConstMod}, the posterior predictive distribution is obtained from the MCMC samples (Sect.~\ref{sec:Mcmc}), while we use the posterior predictive distribution obtained from the spatial interpolation for \mbox{SpatBHM} (Sect.~\ref{sec:Kriging}).
As LocMod is similarly estimated by MCMC, we calculate CRPS once for each element of the LocMod posterior samples, in order to obtain an estimate of the uncertainty of CRPS at each station.
Both the posterior distribution of the constant coefficients in \mbox{ConstMod} and \mbox{SpatBHM} and the spatially interpolated distribution from \mbox{SpatBHM} are assumed as Gaussian, so that we evaluate CRPS following Eq. \ref{eq:CRPSGauss}.
The spatially constant coefficients in \mbox{SpatBHM} were sampled from their posterior distribution for this part of the evaluation, as are their counterparts in \mbox{LocMod}.

The resulting score values are displayed in Fig. \ref{fig:crps} for the six regression coefficients. As $\mu^z$ was not estimated locally, the altitude predictor contribution $\mu^{3}\Delta z_i$ in \mbox{ConstMod} and \mbox{SpatBHM} is added to the intercept for the location parameter, i.e. $\mu^0$, for each station.
The CRPS values show a general improvement of the parameter prediction in \mbox{SpatBHM} compared to \mbox{ConstMod}. The superiority of \mbox{SpatBHM} is most pronounced for $\mu^1$ and $\mu^2$ and less for the scale coefficients. For the $\varsigma^2$ there are hardly any differences. 
From these values, we calculated the skill scores via Eq. (\ref{eq:SkillScore}), shown in Fig. \ref{fig:crpss:ess}a. The CRPS values translate into skill values of up to 50 \unit{\%} for $\mu^2$ of \mbox{SpatBHM} over \mbox{ConstMod}. As $\mu^2$ is one of the GRFs, this clearly points to the added value of the spatial approach.

Following from this, we evaluate the multivariate posterior distribution of the regression coefficients using energy score (Eq. (\ref{eq:EnergyScore})). As can be seen in Fig. \ref{fig:crpss:ess}b, ES shows similarly high skill values of \mbox{SpatBHM} against \mbox{ConstMod}, likewise found around 50 \unit{\%}. In the multivariate case, the skill is rather homogeneous over all stations, pointing to a clear superiority of \mbox{SpatBHM} with regard to the spatial prediction of the parameters.

\begin{figure}
    \centering
    \includegraphics[width=\linewidth]{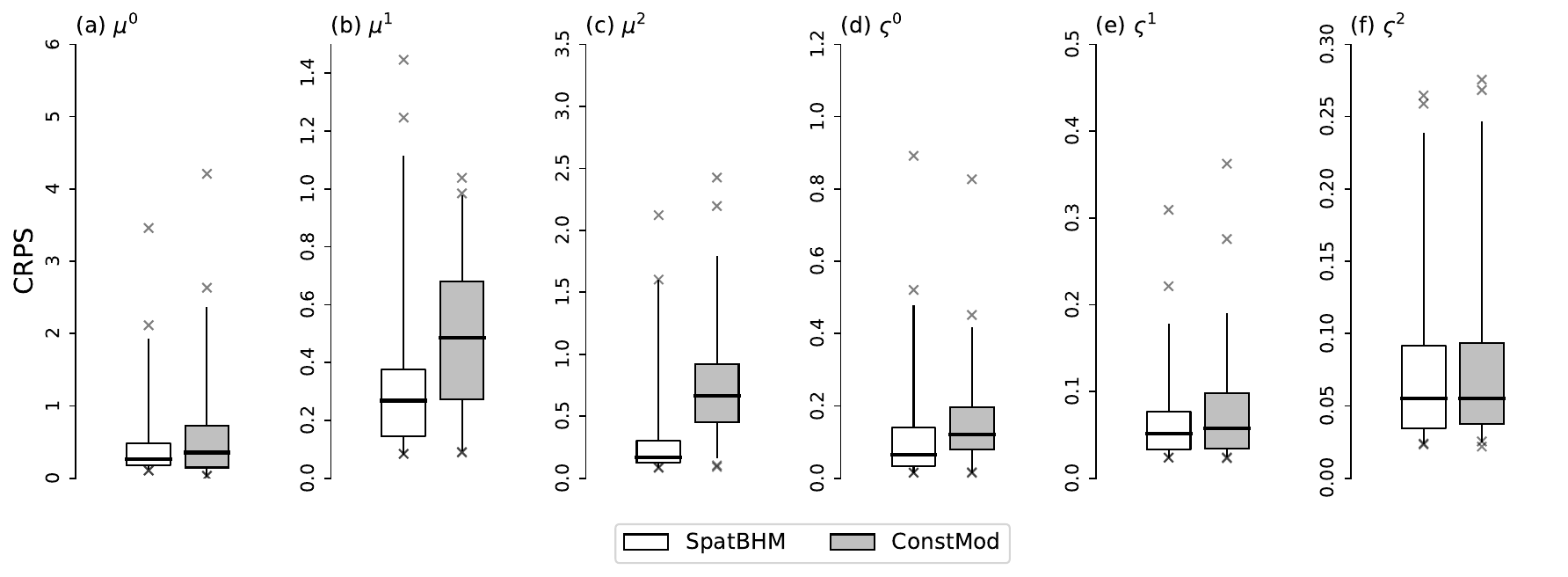}
    \caption{CRPS for the regression coefficients evaluated against the parameter estimations from \mbox{LocMod} fit, trained only on the local data. Score values are shown for (a) $\mu^0$ including the altitude predictor and $\mu^z$, (b) $\mu^1$, (c) $\mu^2$, (d) $\varsigma^0$, (e) $\varsigma^1$ and (f) $\varsigma^2$. Boxplot specifications as in Fig. \ref{fig:SkillConstClimLocal}.}
    \label{fig:crps}
\end{figure}

\begin{figure}
    \centering
    \includegraphics[width=\linewidth]{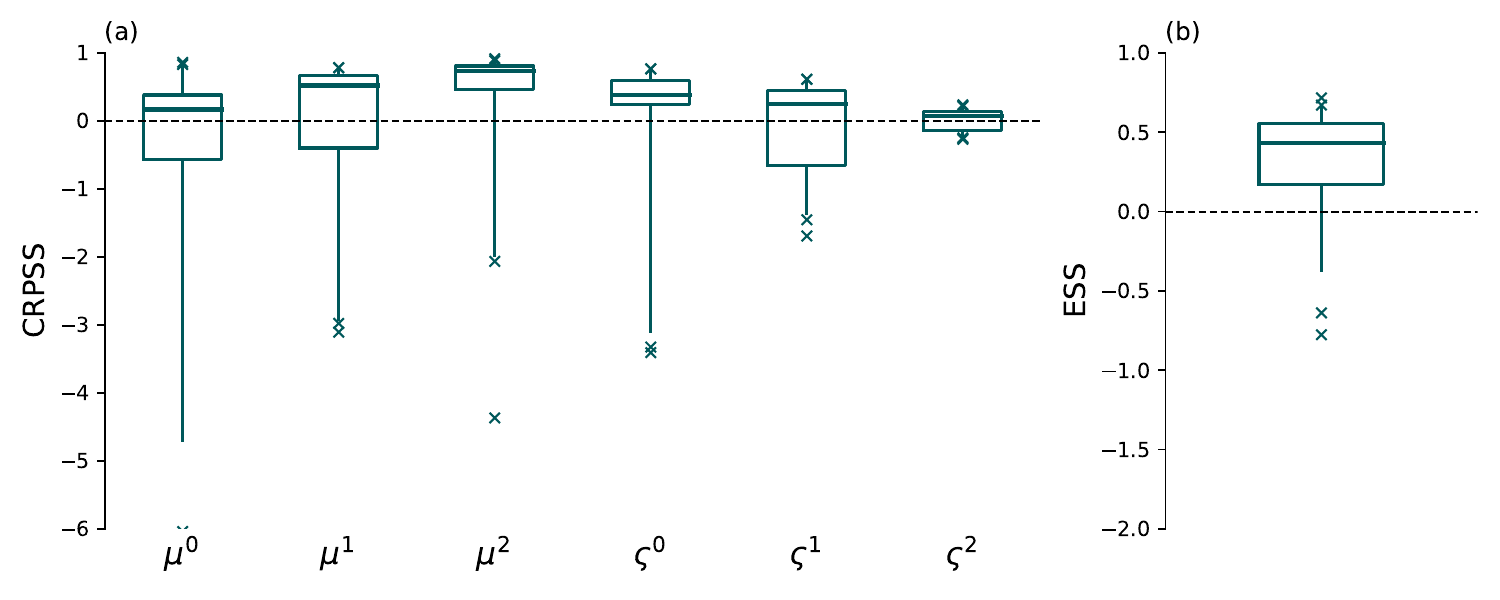}
    \caption{(a) CRPS skill of \mbox{SpatBHM} compared to \mbox{ConstMod} for the regression coefficients. (b) Energy skill score of \mbox{SpatBHM} compared to \mbox{ConstMod} for the multivariate estimation of the regression coefficients. The single parameters are weighted with their spatial variance before multivariate aggregation. Boxplot specifications as in Fig. \ref{fig:SkillConstClimLocal}. }
    \label{fig:crpss:ess}
\end{figure}

\subsection{Post-processing time series}
\label{sec:TimeSeries}

\begin{figure}[ht]
    \centering
    \includegraphics[width=\textwidth]{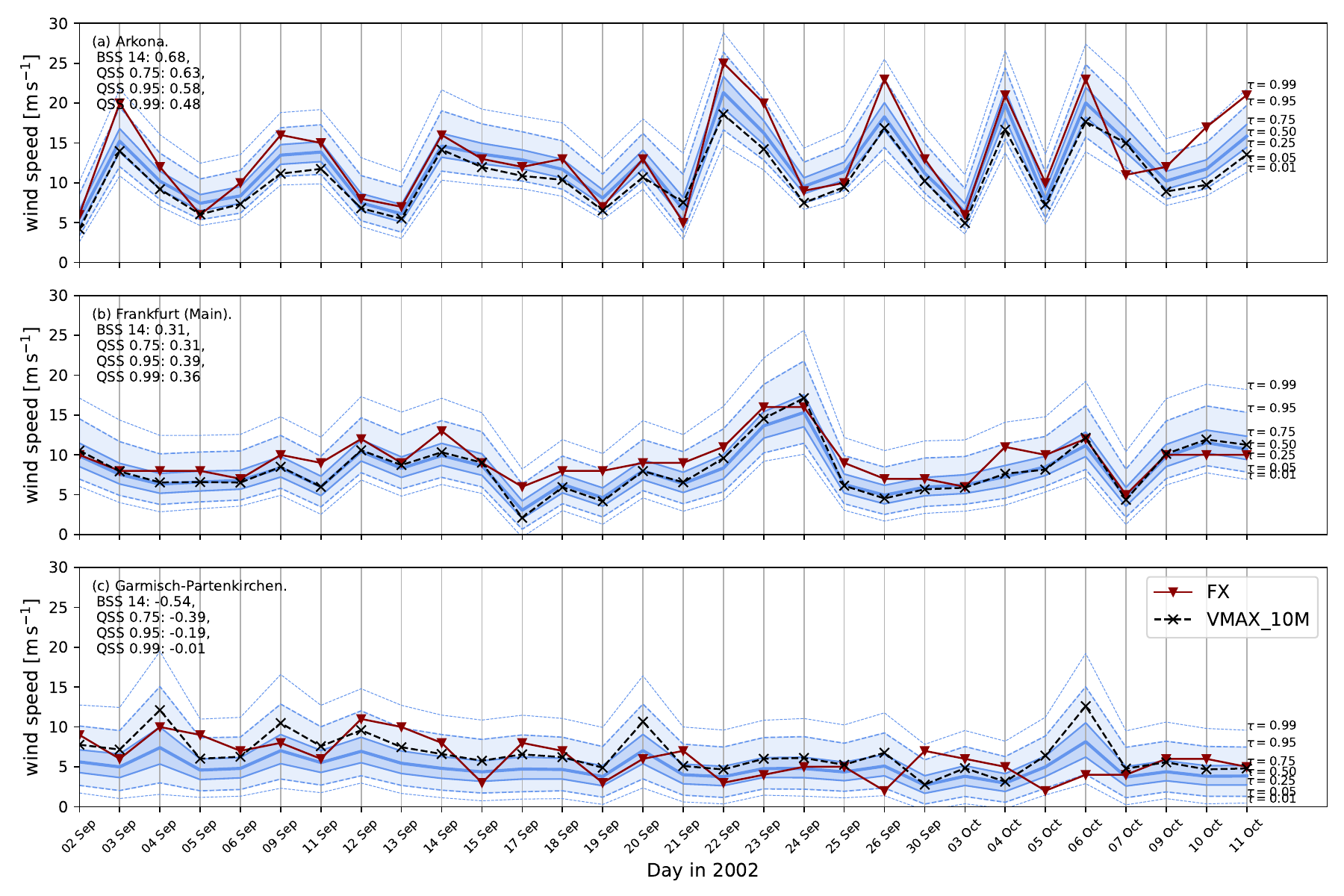}
    \caption{Predictive distributions generated by \mbox{SpatBHM} in 2002 for (a) Arkona, (b) Frankfurt (Main) and (c) Garmisch-Partenkirchen. The station observations $\mathrm{FX}$ are marked by the red line, while $V_\mathrm{max}$ is marked in black. The blue shading denotes the various quantiles of the predictive distribution, as given by the annotations. The displayed skill score values are calculated against climatology. It should be noted that the time series are incomplete due to missing observations. The predictions are cross-validated, i.e. each station was removed from the data prior to model training and prediction.}
    \label{fig:PredSpatbhm}
\end{figure}

The objective of the spatial hierarchical extreme value model is to provide an enhanced description of surface wind gusts in reanalysis by providing a probabilistic diagnostic for wind gusts based on a statistical post-processing of predictor variables.
\mbox{SpatBHM} stands to the task quite well as illustrated by Fig.~\ref{fig:PredSpatbhm}, depicting time series of postprocessed wind gust from \mbox{SpatBHM} at three distinct locations for September and October 2002. 
The observations consistently fall within the displayed distributional range. 
Moreover, at the first station, the median of the predicted distributions is visibly corrected towards the corresponding observations, indicating that the statistical model improves the calibration of the reanalysis.
The third location in Fig.~\ref{fig:PredSpatbhm}c is used to contrast the results with a location with no skill.

Figure~\ref{fig:PredSpatbhm}a depicts the postprocessed wind gusts at Arkona, a coastal station in northeastern Germany. 
The post-processing is capable of capturing the transient weather patterns, which is especially evident from 21 September to 7 October. 
The skill score values against climatology are quite high with values larger than 50 \unit{\%}, and the observations typically lie inside the 98 \unit{\%} confidence interval of the predicted distributions.
The forecasts are well calibrated at this location, as can be understood from the probability integral transform (PIT) histogram \citep{Dawid84} in Fig. \ref{fig:PITStations}a for the same station. In general, the histogram looks homogeneous with an indication of a too high dispersion of the forecast distribution. 

For central German locations, \mbox{SpatBHM} shows a similarly strong performance, as illustrated by the example of Frankfurt in Fig.~\ref{fig:PredSpatbhm}b. In most cases, the observed wind gusts are found close to the median of the predicted distribution.
However, $V_{\mathrm{max}}$ and the observations display usually similar values at this location, so that the surface maximum wind speed in the reanalysis can already be regarded as a good prediction for the observation.
At this location, the PIT-histogram (Fig. \ref{fig:PITStations}b) provides interesting added information on the gust prediction. Apparently, there is a consistent negative bias, combined with a too large variance of the forecast distribution. This suggests that the high skill is achieved by a bias-variance trade-off, where a larger variance compensates the negative bias for the quantiles and thresholds of interest.

The spatial post-processing encounters challenges in complex terrain with low wind speed, as illustrated by the example of Garmisch-Partenkirchen in Fig.~\ref{fig:PredSpatbhm}c.
This station is the location with the lowest average wind speed among the selected stations and even \mbox{LocMod} has no skill.
At this location, \mbox{SpatBHM} corrects the observations systematically to lower values, which leads to a pronounced negative bias.
Accordingly, the PIT-histogram in Fig. \ref{fig:PITStations}c shows a similarly strong negative bias. Different to Fig. \ref{fig:PITStations}b, there is no visible excess of dispersion of the forecast. Therefore, the bias-variance trade-off does not succeed at this location.
However, we still observe a slight improvement in skill with respect to \mbox{ConstMod}.
The low skill at this location is due to the lack of explanatory power of the predictor variables at this particular location, as shown by the non-systematic bias and a low correlation between the reanalysis and the observations.

\begin{figure}
    \centering
    \includegraphics[width=\linewidth]{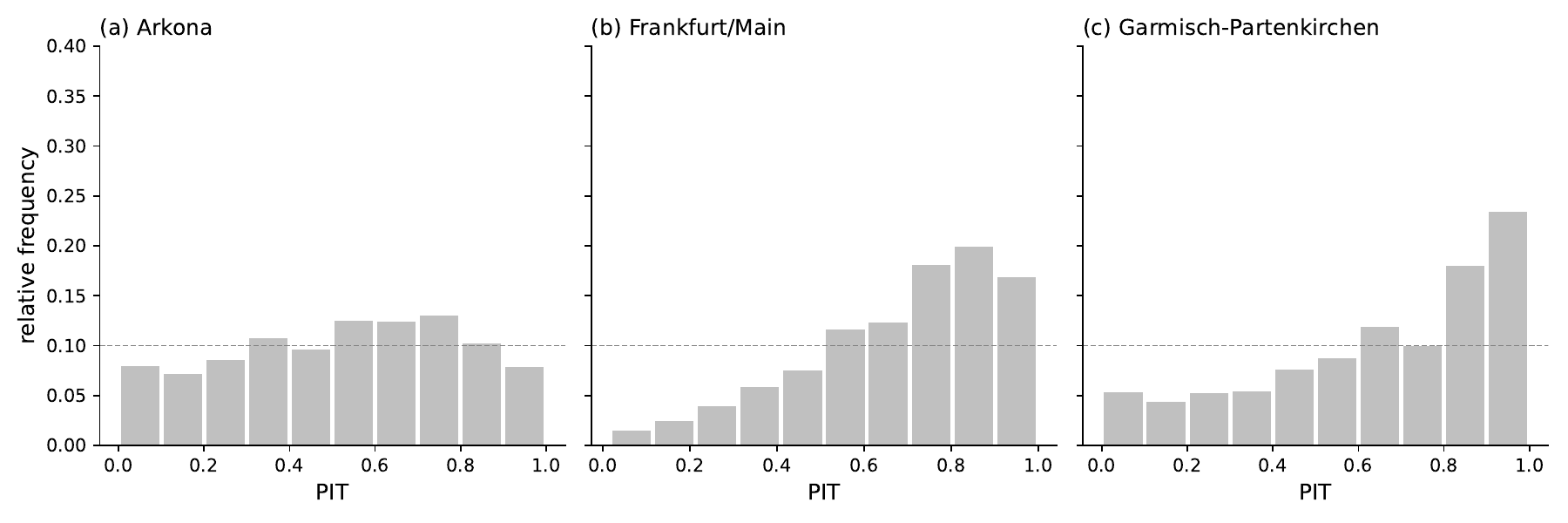}
    \caption{PIT histograms for the three stations from Fig. \ref{fig:PredSpatbhm}. PIT-values are obtained by evaluating the predicted empirical gust distribution function at the observed values.}
    \label{fig:PITStations}
\end{figure}

\subsection{Post-processing on a spatial grid}
\label{sec:GridInterpolation} 

Post-processing gridded data sets using \mbox{SpatBHM} involves the application of the spatial interpolation procedure to high dimensional data. In theory, this is readily doable because of the use of GRFs in the model formulation.
However, the interpolation process involves the inversion of the covariance matrix, which is computationally expensive in high dimensions. Therefore, we advise to split the target region in subregions and to draw iteratively from the GRF. We start by simulating a horizontal slice of the data set in the south and iteratively simulate horizontal slices further northward.
In each iteration, the probability for the next horizontal slice is conditioned on the results of the previous draw for the adjoining slice in addition to conditioning on the fitted values at the training locations.
This iterative drawing process is repeated, until the complete field has been simulated.
In the absence of a unique station altitude for each grid cell, we set the $\Delta z$ predictor to zero. Thereby, we enable a statistical downscaling of the gust model to the subgrid-scale in the event that the altitude predictor exists.
\begin{figure}
    \centering
    \includegraphics[width=\linewidth]{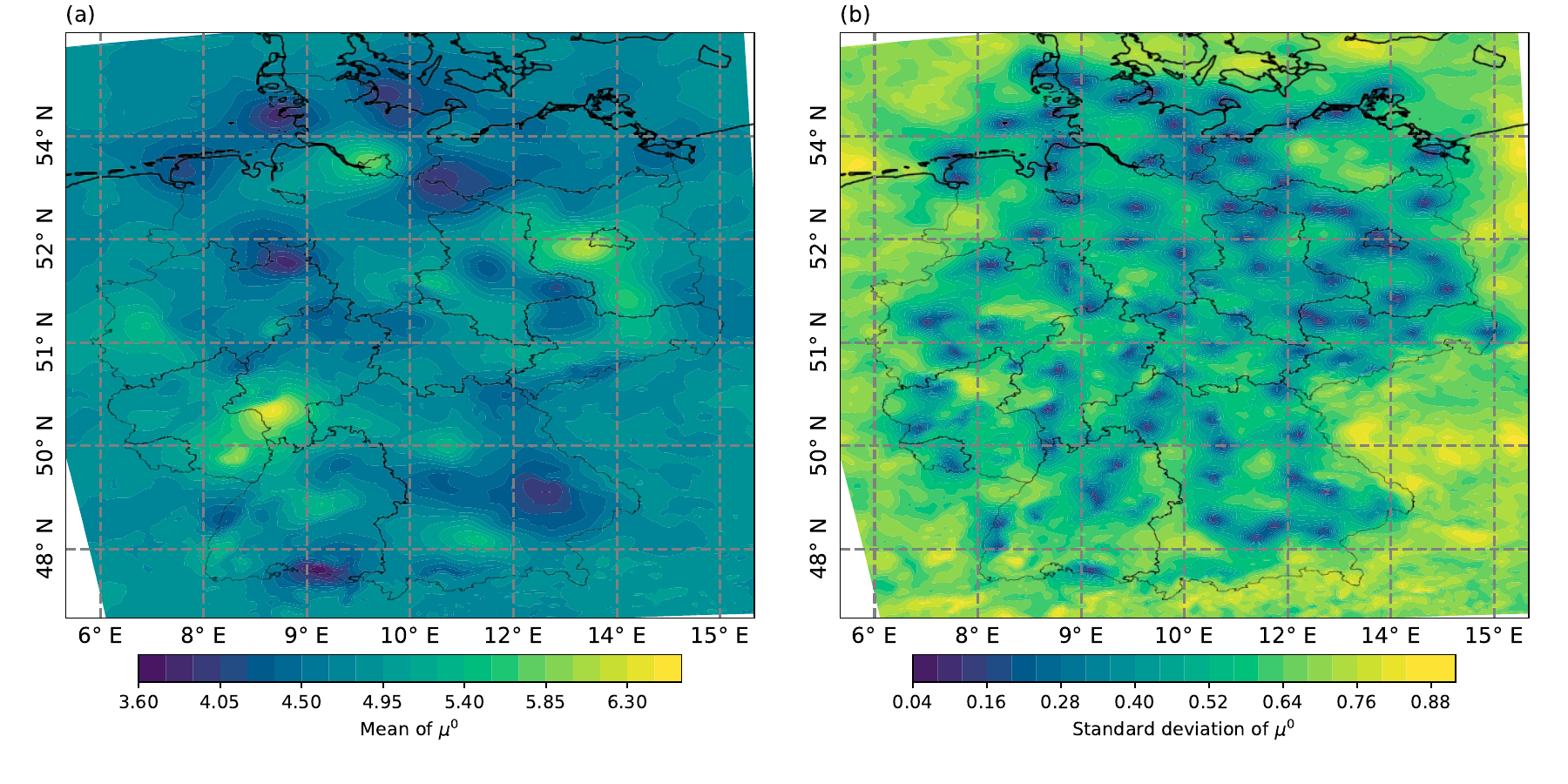}
    \caption{Posterior distribution of $\mu_0$ after spatial interpolation. Left: expectation for the coefficient $\mu_0$, interpolated from \mbox{SpatBHM}. Right: The standard deviation of the same field based on 100 draws from the same model.}
    \label{fig:GRF:Interpolated}
\end{figure}

Figure~\ref{fig:GRF:Interpolated}a provides an example for the result of this iterative interpolation, obtained for $\mu_0$, for a grid of 121 times 161 cells over Germany. Each simulated horizontal slice contains 5 rows of grid cells, resulting in 605 simultaneously drawn representations in each iteration.
The mean of $\mu_0$ is a smooth spatial field with some visible minima and maxima, not all of which are easily explainable by the geography. However, some geographical features, such as the Hartz Mountains in the center and the Cologne Lowland in the west are discernible. The parameter predictions are lower over the ocean and the adjoining coastlines. Areas that are sufficiently far away from any of the training locations, are estimated close to the spatial field mean. Owing to the small range parameter estimates (cf. Table~\ref{tab:SpatbhmParvalues}), the process is still considerably rough, as it represents a local station correction to the spatial mean and \mbox{SpatBHM} does not include a nugget effect. A nugget effect could account for outlying stations, but we did not include it as it decreased the predictive performance in preliminary tests.
Figure~\ref{fig:GRF:Interpolated}b shows the standard deviation of the predicted $\mu_0$-field, obtained from 100 draws of the same spatial field. The variability of the interpolated field is lower near the training locations and increases with distance. Mountain stations, such as Zugspitze, Brocken and Feldberg are not as distinctly visible as non-mountain stations, as the elevation offset (Eq. \ref{eq:AltitudeEuclidean}) successfully segregates them from their immediate surroundings. Thereby, their influence on the direct neighborhood is reduced, albeit they remain inside the training data set.

\begin{figure}
    \centering
    \includegraphics[width=\linewidth]{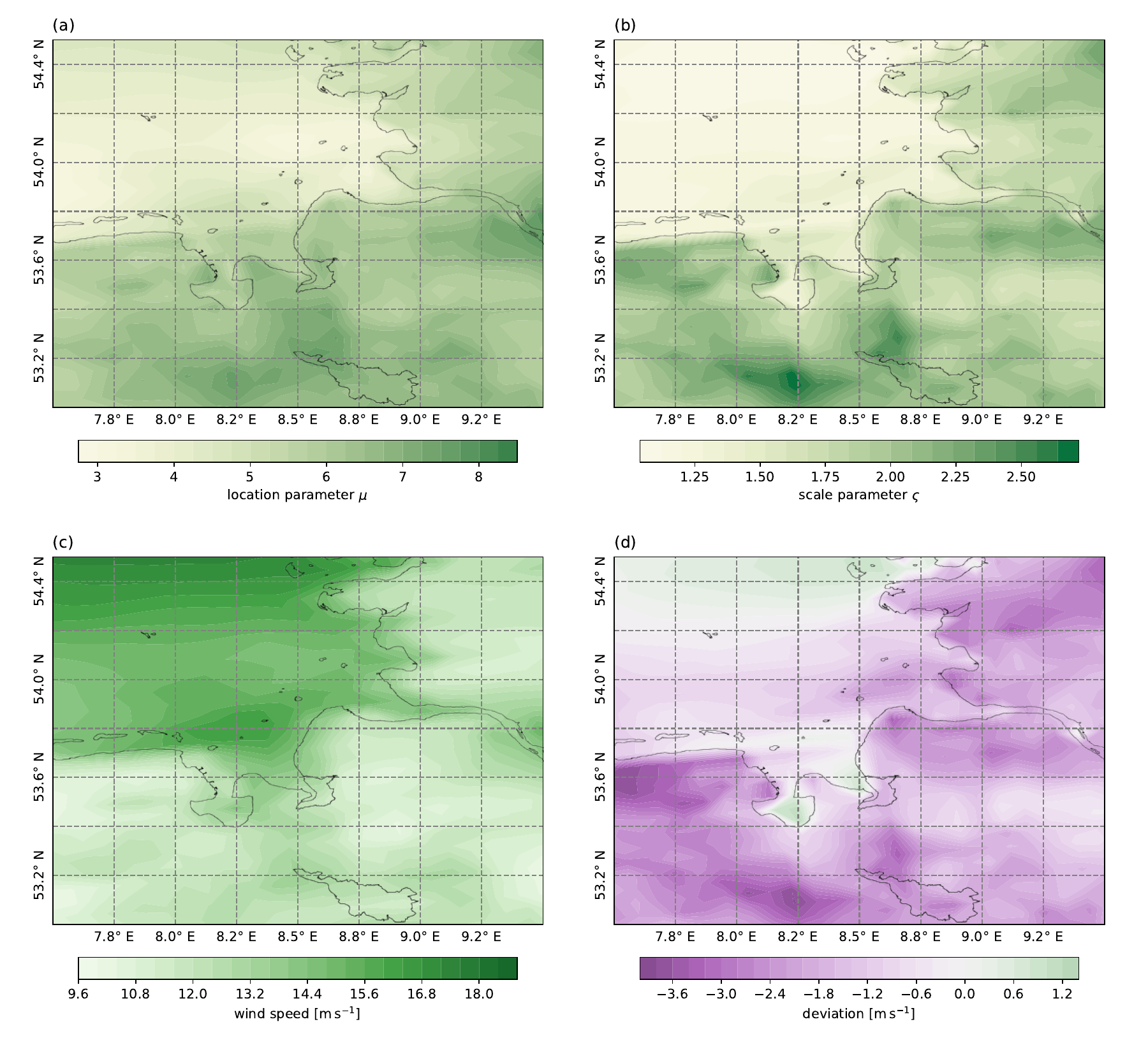}
    \caption{Spatial fields of (a) the location parameter and (b) the scale parameter on 15 September 2010, predicted by \mbox{SpatBHM} over the German North Sea area. (c) Postprocessed expectation value for $\mathrm{FX}$ from SpatBHM. (d) Difference between postprocessed expectation value and the $V_\mathrm{max}$ predictor field. Green shading indicates a positive deviation of SpatBHM and purple indicates a negative deviation.}
    \label{fig:SpatialPrediction}
\end{figure}
Following the interpolation of the GRFs, it is possible to estimate spatial fields of the Gumbel parameters of the wind gust distribution.
Figure~\ref{fig:SpatialPrediction}a and b depict such spatial fields for the location and the scale parameter, including the predictor variables. The shown area is a portion of the German North Sea coastline.
The fields are obtained from \mbox{SpatBHM} by retrieving one representation of the interpolated spatial fields for the coefficients $\mu^{0}$ and $\mu^{2}$ and evaluating \mbox{Eq. (\ref{eq:ConstMod})} for the predictor variables on 15 September 2010, while setting $\Delta z_i \equiv0$.\\
There is a strong contrast between the sea and land surfaces due to the contribution of the covariate variables. Both location and scale parameter are estimated higher over land than over the ocean. For the scale parameter, this contrast is even more pronounced. Apart from the land-sea-difference, the fields look reasonably smooth. 
Although the higher parameter predictions over land appear counterintuitive at first, they are plausible considering that \mbox{SpatBHM} does not predict the gust speed, but $\mathrm{FX}-V_\mathrm{m}$. Adding $V_\mathrm{m}$ to the expectation value of the \mbox{SpatBHM} prediction results in overall higher postprocessed gust predictions over the ocean than over land as shown, in Fig.~\ref{fig:SpatialPrediction}c. Concluding from this, wind gusts possess considerably less variability over the ocean and are found closer to the mean wind values. Therefore, the model parameters are estimated lower.
Figure \ref{fig:SpatialPrediction}d shows the difference between the $V_\mathrm{max}$ field from COSMO-REA6 and the expectation value of the postprocessed field from \mbox{SpatBHM}. In most areas, the correction applied by SpatBHM to the expectation is negative, i.e. reducing the mean wind prediction. Only two regions display a positive correction, namely the northern depicted part of the ocean and two intricate bays (Jade bight, Weser estuary), which are standing out. This difference compared to the surrounding area comes from the interpolation of land-area location parameters, that are higher than location parameters over sea. Simultaneously, $V_\mathrm{m}$ is higher over the bay areas due to reduced roughness. The combination of these two factors leads to the distinctly positive correction. However, as we do not have any verifying observations over the ocean we refrain from making any statements about the accuracy of this correction. Nevertheless, including the land-sea-mask as spatial predictor can be a useful addition for \mbox{SpatBHM} in the future.\\

\clearpage
\conclusions  
\label{sec:Conclusions}
In this study, we present a spatial probabilistic post-processing approach (\mbox{SpatBHM}) for wind gusts from reanalysis models, with the COSMO-REA6 model serving as a case study.
The probabilistic post-processing is based on Bayesian hierarchical modeling and extreme value statistics.
Wind gust observations are modeled as a non-stationary Gumbel distribution, with location and scale parameters varying both spatially and temporally. 
The temporal dependency is modeled as a linear model on spatio-temporal predictor variables. The spatial dependency is included by letting the regression coefficients vary in space following a Gaussian random field with an isotropic covariance structure. 
We compared the \mbox{SpatBHM} against a model with spatially constant coefficients (\mbox{ConstMod}) using proper scoring rules for quantiles and threshold exceedance probabilities and their decomposition to assess miscalibration, resolution, and uncertainty.

Although we used a computationally expensive MCMC-algorithm for model fitting, the sampling process can be adapted for operational use. Pystan (3.9.0) with Python (3.9.18) was run sequentially in a JupyterHub service running Debian GNU/Linux 12 (x86\_64), requiring about one hour of sampling time for the final version of SpatBHM. Training can be sped up by about 25~\% by selecting a more informative prior for the range parameters.
The spatial interpolation to large data sets is a computational bottleneck because the parameter fields must be interpolated to the entire domain for all elements of the posterior sample. Thus, the entire process must be repeated $N$ times. For this analysis, the process was carried in R (4.2.2) on a computer with an AMD Ryzen 2400G processor, running at 3.6 GHz using 29 GB of RAM, and an openSUSE Leap 15.5 system (x86\_64). The process described in Sect.~\ref{sec:GridInterpolation} with $N=100$ consumed about 3h of computation time, so the complete analysis can be reproduced on a standard desktop ($\geq8$ cores, 16 GB RAM). For operational purposes, the spatial interpolation can also be run in parallel.

We have reached a proof of concept in favor of the spatial modeling approach, indicating that \mbox{SpatBHM} possesses higher skill than \mbox{ConstMod} for all investigated features of the predictive distribution, although we find that ConstMod already exhibits high skill against the climatology.
While \mbox{SpatBHM} does not show much improvement for the prediction of threshold exceedance probabilities, it clearly outperforms \mbox{ConstMod} for prediction quantiles by 2--5 \unit{\%} in terms of median skill, depending on the exact quantile under investigation. 
Moreover, SpatBHM demonstrates particular skill in predicting high quantiles, due to a better representation of the variability of wind gusts.
This is evident more when focusing the evaluation on top quartile of the predictions.
As a further advantage, the GRFs in \mbox{SpatBHM} can be readily interpolated to large data sets by an iterative simulation procedure. Therefore, the model facilitates simple post-processing including gridded data sets.

However, the skill of all models is low at locations with little wind.
Provided that a contribution of the station altitude difference is included into the estimation of the spatial covariance structure, \mbox{SpatBHM} can enhance the predictive skill at these locations to some degree.
The elevation offset stabilizes the estimation of the range parameter of the GRF and results in a more accurate and better calibrated representation of the spatial structure.
However, even the spatial model falls short of attaining the same level of calibration as the simple climatology in these low wind conditions. 
An examination of the score decomposition in these cases revealed that the challenges observed at these locations are not solely caused by poor calibration, but also by a lack of model resolution. 
This is an indication of a lack of explanatory power of the predictor variables in low-wind conditions.

In conclusion, wind gust post-processing remains a challenging topic due to the localized nature, short duration, and inherent extremity of wind gusts.
Further research is required to account for an annual or diurnal cycle and to assess the models' performance in different boundary layer regimes.
One way to include the annual and diurnal cycles is to use harmonic functions with suitable frequencies as additional predictors. Further testing is required to determine whether their effects are best modeled as spatially constant or as GRFs. Currently, the model code allows the flexible inclusion of new predictors by adding columns to the predictor matrix and specifying whether their coefficient is modeled as constant or as a GRF. The spatial interpolation process remains unchanged, but the prediction code requires updates to include the harmonic function predictors. In summary, the annual and diurnal cycles can be easily included, but their inclusion requires further model checking and comparison, as well as a more extensive data preprocessing.
Finally, developing a spatial prediction procedure, that is capable of accounting for existing gust observations at specific locations, namely, conditionally sampling from the complete gust model, can be an interesting follow up to this work.
As the spatial post-processing approach demonstrates added skill to existing linear approaches, it can contribute to an improved representation of wind gust characteristics in reanalysis.




\codedataavailability{The SYNOP observations and COSMO-REA6 reanalysis data used in this study are publicly accessible through DWD (German weather service) via an open data server (https://opendata.dwd.de). The model training was performed using Stan (https://mc-stan.org), using the Python interface pystan. The spatial interpolation and prediction were implemented using GNU-licensed free software from the R Project for Statistical Computing (http://www.r-project.org). The Stan model code is provided along a Jupyter Notebook with model training examples, the preprocessed training and evaluation data sets and the R-Code for the spatial interpolation and the prediction are available from Zenodo at https://doi.org/10.5281/zenodo.15437958 \citep{Ertz25code}. The same repository contains two recreation Jupyter Notebooks for test experiments regarding the shape parameter of the GEV distribution and the positive definiteness of the covariance matrix, that have been suggested by the associate editor and one anonymous reviewer.} 



\appendix
\section{List of symbols}\label{A:symbols}
\appendixtables
\begin{table}
    \caption{Notation used throughout this manuscript. }
    \label{tab:notation}
    \centering
    \begin{tabular}{l|l}
         $\pi$ & probability density function \\
         $y_{it}$ & observation at location $\vec{r}_i$, $i=1, \ldots, q$ and time $t_k$, $k=1, \ldots,n$ \\
         $\mu_{it}$, $\varsigma_{it}$ & GEV parameters at location $\vec{r}_i$, $i=1, \ldots, q$ and time $t_k$, $k=1, \ldots,n$\\
         $x^\mu_{j,ik}$/ $x^\varsigma_{j,ik}$ & set of covariates $j=1, \ldots,m_\mu$ for $\mu_{it}$ / $j=1, \ldots,m_\varsigma$ for $\varsigma_{it}$ \\
         $\vec{x}^\mu_{it}$/ $\vec{x}^\varsigma_{it}$ & vector of covariates  $\vec{x}^\mu_{it} = (x^\mu_{0,ik}, \ldots, x^\mu_{m_\mu,ik})$  / $\vec{x}^\varsigma_{it} = (x^\varsigma_{0,ik}, \ldots, x^\varsigma_{m_\varsigma,ik})$\\
         $\mu^j_i$ / $\varsigma^j_i$ & time constant regression coefficients for covariate $x^\mu_{j,ik}$/ $x^\varsigma_{j,ik}$ \\
         $\vec{\beta}^\mu_i$ / $\vec{\beta}^\varsigma_i$ & vector with regression coefficients $\vec{\beta}^\mu_i = (\mu^{0}_{i}, \ldots, \mu^{{m_\mu}}_{i})$ /$\vec{\beta}^\varsigma_i = (\varsigma^{0}_{i}, \ldots, \varsigma^{{m_\varsigma}}_{i})$\\
         $\vec{\mu}^{j}$ / $\vec{\varsigma}^{j}$ & $\vec{\mu}^{j} = (\mu^{j}_{1}, \ldots,\mu^{j}_{q})$ / $\vec{\varsigma}^{j} = (\varsigma^{j}_{1}, \ldots,\varsigma^{j}_{q})$ \\
         $\vec{\alpha}_{\mu^{j}}$, $\boldsymbol{\mathrm{\Sigma}}_{\mu^{j}}$ / $\vec{\alpha}_{\varsigma^{j}}$, $\boldsymbol{\mathrm{\Sigma}}_{\varsigma^{j}}$ & parameters of multivariate Gaussian distribution for regression coefficients $\vec{\mu}^{j}$ / $\vec{\varsigma}^{j}$ \\
         $\alpha_{\mu^{j}}, \sigma_{\mu^{j}}, \rho_{\mu^{j}}$ / $\alpha_{\varsigma^{j}}, \sigma_{\varsigma^{j}}, \rho_{\varsigma^{j}}$ & parameters of homogeneous GRFs for regression coefficients $\mu^{j}(\vec{r})$ / $\varsigma^{j}(\vec{r})$\\
         $\gamma_{\mu^{j}}, \tau^2_{\mu^{j}},\delta_{\mu^{j}}, \epsilon_{\mu^{j}}\zeta_{\mu^{j}}, \eta_{\mu^{j}} $& parameters of priors for random field parameters of $\mu^{j}(\vec{r})$\\
         $\gamma_{\varsigma^{j}}, \tau^2_{\varsigma^{j}},\delta_{\varsigma^{j}}, \epsilon_{\varsigma^{j}}\zeta_{\varsigma^{j}}, \eta_{\varsigma^{j}} $& parameters of priors for random field parameters of $\varsigma^{j}(\vec{r})$\\
         $f_z$ & scaling factor for elevation offset
    \end{tabular}
\end{table}

\clearpage
\section{Predictor selection for linear models}\label{A:PredictorSelection}
The selection of the optimal predictor variables for the spatial hierarchical post-processing model was performed on  \mbox{ConstMod}, which also serves as baseline model during evaluation in Sect. \ref{sec:ConstMod}.
We tested and evaluated a variety combinations of predictors and predictands.
For \mbox{ConstMod}, all regression coefficients $\vec{\mu}^{j}$ and $\vec{\varsigma}^{j}$ are spatially constant.
All model versions are trained as outlined in Sect. \ref{sec:ConstMod}.
Table~\ref{tab:ConstmodOverview} provides an overview over the all trained versions, including the selection of predictand and predictor variables for location and scale, as well as the median score values obtained at the 109 verification locations.
The predictor variables follow the nomenclature as outlined in Sect. \ref{sec:ConstMod}.
The locally estimated model local model (\mbox{LocMod}), uses the same formulation as \mbox{ConstMod~4} without the altitude predictor $\Delta z$. By estimating \mbox{ConstMod} locally for each station, the contribution of $\Delta z$ is directly added to the intercept $\mu^{0}$.
\mbox{LocMod} is trained on the local station data only, so it learns the local wind gust characteristics.
For purposes of comparison, an overview of all \mbox{ConstMod} versions, LocMod and the climatology are presented in Table~\ref{tab:ConstmodOverview} together with the median score values for assessment of the predictive skill.
Figure~\ref{fig:SkillConstmod} shows the cross-validated skill scores for \mbox{ConstMod} compared to the local climatology for the 14 and 18 \unit{m~s^{-1}} threshold and the $0.75$, $0.95$, $0.99$ and $0.999$-quantiles for the different versions of \mbox{ConstMod}, obtained at all investigated locations. 
All versions of \mbox{ConstMod} demonstrate a high level of skill against the local climatology. The skill is large over all investigated distributional features with average skill score values between 10 \unit{\%} and 45 \unit{\%}.

\begin{figure}
    \centering
    \includegraphics[width=\textwidth]{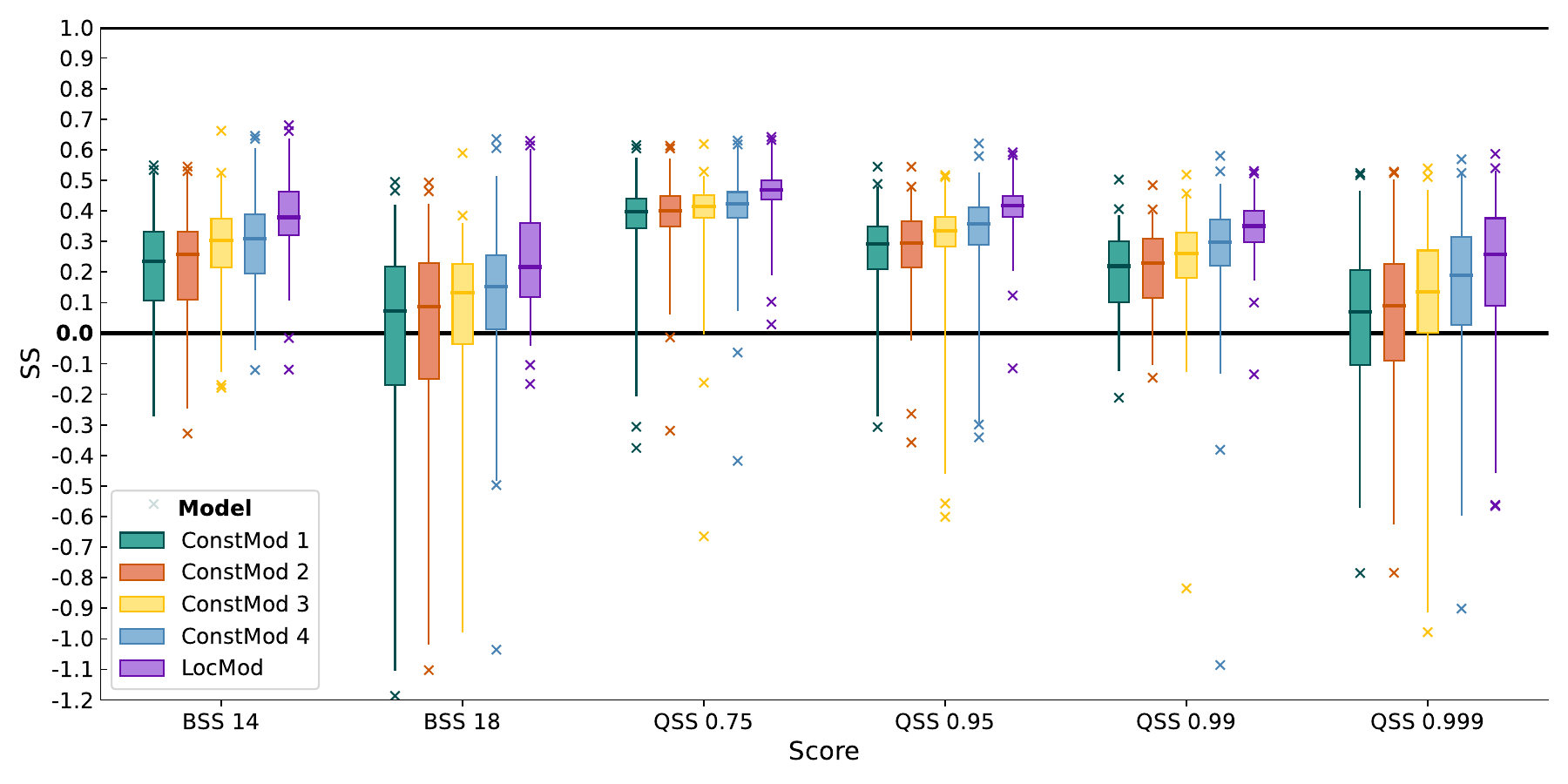}
    \caption{Cross-validated skill scores against climatology of various versions of \mbox{ConstMod} and LocMod (colored boxes) for exceedance probabilities of the 14 \unit{m~s^{-1}} (BSS 14) and 18 \unit{m~s^{-1}} (BSS 18) threshold, and the $0.75$ to $0.999$-quantiles (QSS 0.75 to QSS 0.999). Each box plot contains skill scores at 109 locations. Boxes represent the interquartile range and whiskers extend to the $0.01$ and $0.99$-quantiles. Bold lines mark the median scores. Model definitions and median scores are given in Table~\ref{tab:ConstmodOverview}.}
    \label{fig:SkillConstmod}
\end{figure}

The model version exhibiting the least skill is \mbox{ConstMod 1}, which only uses $V_{\mathrm{max}}$ as predictor.
The incorporation of the altitude predictor $\Delta z$ (\mbox{ConstMod~2}) does not improve the skill for non-mountain stations, but it is visible that the quantile predictions at some locations are improved compared to the simplest model version. 
The first noticeable improvement in skill is observed when the predictand is modified to $\mathrm{FX}-V_\mathrm{m}$ (\mbox{ConstMod~3}). This finding is consistent with the findings of \citet{Friederichs09}.
The improvement in skill is particularly evident for the 14 \unit{m~s^{-1}} threshold and for higher quantiles.
The optimal spatially constant linear model was found to be the full version (\mbox{ConstMod 4}), as specified in Eq.~(\ref{eq:ConstMod}).
Therefore, we base all spatial models on \mbox{ConstMod 4}.
The final fitted values of the parameters for the various ConstMod versions are presented in Table~\ref{tab:ConstmodParvalues}.

\begin{table}[ht]
    \caption{Model versions and median score values for threshold excess probabilities and quantiles. For climatology, no predictors are used and parameters are estimated at each station, respectively. \mbox{ConstMod} 1--4 are estimated at all 109 stations simultaneously. \mbox{LocMod} refers to a local version of \mbox{ConstMod 4} without $\Delta z$  trained at each location separately.}
    \label{tab:ConstmodOverview}
    \centering
    \begin{tabular}{c|ccc|cccccc}
    \tophline
    \textbf{Model} & \textbf{Predictand} & \textbf{Predictors $\mu$} & \textbf{Predictors $\varsigma$} & BS 14 & BS 18 & QS 0.75 & QS 0.95 & QS 0.99 & QS 0.999\\
    \middlehline
    Climatology &  $\mathrm{FX}$ & none & none & 0.0637 & 0.0113 & 1.105 & 0.414 & 0.120 & 0.0163 \\
    ConstMod 1 & $\mathrm{FX}$ & $V_\mathrm{max}$ & $V_\mathrm{max}$ & 0.0479 & 0.0113 & 0.666 & 0.297 & 0.094 & 0.0144\\
    ConstMod 2 & $\mathrm{FX}$ & $V_\mathrm{max}$, $\Delta z$ & $V_{\mathrm{max}}$ & 0.0477 & 0.0113 & 0.659 & 0.295 & 0.093 & 0.0142 \\
    ConstMod 3 & $\mathrm{FX}-V_\mathrm{m}$ & $V_\mathrm{max}$, $\Delta z$ & $V_\mathrm{max}$ & 0.0443 & 0.0103 & 0.639 & 0.271 & 0.086 & 0.0134\\
    ConstMod 4 & $\mathrm{FX}-V_\mathrm{m}$ & $V_\mathrm{max}$, $V_\mathrm{m}$, $\Delta z$ & $V_{\mathrm{max}}$, $V_\mathrm{m}$ & 0.0433 & 0.0103 & 0.636 & 0.265 & 0.084 & 0.0127 \\
    LocMod & $\mathrm{FX}-V_\mathrm{m}$ & $V_\mathrm{max}$, $V_\mathrm{m}$ & $V_\mathrm{max}$, $V_\mathrm{m}$& 0.0397 & 0.0092 & 0.597 & 0.244 & 0.077 & 0.012 \\
    \bottomhline
    \end{tabular}
\end{table}

\begin{table}
    \caption{Parameter estimations for all \mbox{ConstMod} version together with their 99 \unit{\%} confidence intervals, assuming normal marginal posterior distributions. The values are obtained from training the model on the complete traning data set.}
    \label{tab:ConstmodParvalues}
    \centering
    \begin{tabular}{c|cccc}
    \tophline
    \textbf{Parameter} & \textbf{ConstMod 1} & \textbf{ConstMod 2} & \textbf{ConstMod 3} & \textbf{ConstMod 4}\\
    \middlehline
    $\mu^0$   & $8.334 \pm 0.016$ & $8.288 \pm 0.016$ & $4.681 \pm 0.014$ & $4.735 \pm 0.014$ \\
    $\mu^1$   & $2.457 \pm 0.016$ & $2.445 \pm 0.016$ & $1.118 \pm 0.014$ & $1.843 \pm 0.027$\\
    $\mu^2$   & $-$ & $-$ & $-$ & $-0.890 \pm 0.025$\\
    $\mu^z$ & $-$ & $0.152 \pm 0.007 $ & $0.221 \pm 0.007 $ & $0.226 \pm 0.008$\\
    $\varsigma^0$ & $0.680 \pm 0.005$ & $0.673 \pm 0.005$ & $0.556 \pm 0.005$ & $0.535 \pm 0.005$\\
    $\varsigma^1$ & $0.330 \pm 0.005$ & $0.332 \pm 0.005$ & $0.218 \pm 0.005$ & $0.403 \pm 0.009$\\
    $\varsigma^2$ & $-$ & $-$ & $-$ &$-0.239 \pm 0.008$\\
    \bottomhline
    \end{tabular}
\end{table}

\section{Selection of spatial fields for SpatBHM}\label{A:SpatialFields}

Subsequent to the implementation of the elevation offset, SpatBHM is constructed in a gradual manner, commencing from ConstMod 4 and permitting an increasing number of model parameters to vary in space. This approach permits the assessment of the added value of individual spatial parameters.
Firstly, only the intercept of the location parameter, $\mu^0$, is made variable in space, while the remaining parameters are kept spatially constant. Subsequently, the number of spatially varying parameters is increased through the testing of different combinations of spatial and non-spatial parameters, as illustrated in Table~\ref{tab:ModelOverviewSpatial}.
The parameter $\mu^z$ (i.e. the regression coefficient for $\Delta z$) is spatially constant in all model versions, as it controls a spatial covariate assuming a fixed value at each location.
The fitted values of the parameters are shown in Table~\ref{tab:SpatbhmParvalues}.
The median score values for \mbox{SpatBHM} are displayed in Table~\ref{tab:ModelOverviewSpatial}.
All spatial models demonstrate a comparable level of skill with respect to climatology as ConstMod across the entire range of distribution characteristics investigated. 

\begin{table}
    \caption{\mbox{SpatBHM} versions with model name, spatially variable regression coefficients and cross-validated median score values obtained from the 109 stations. In each model, the remaining regression coefficients are assumed to be spatially constant.}
    \label{tab:ModelOverviewSpatial}
    \centering
    \begin{tabular}{c|c|cccccc}
    \tophline
    \textbf{Model} & \textbf{Spatial parameters} & BS 14 & BS 18 & QS 0.75 & QS 0.95 & QS 0.99 & QS 0.999\\
    \middlehline
    SpatBHM 1 & $\mu^0$ & 0.0427 & 0.01 & 0.633 & 0.261 & 0.082 & 0.0124 \\
    SpatBHM 2a & $\mu^0$, $\mu^1\,(V_\mathrm{max})$ & 0.0434 & 0.0101 & 0.6308 & 0.260 & 0.081 & 0.0121 \\
    SpatBHM 2b & $\mu^0$, $\mu^2\,(V_\mathrm{m})$ & 0.0435 & 0.01 & 0.6314 & 0.2598 &  0.0805 & 0.0121 \\
    SpatBHM 2c & $\mu^0$, $\sigma^0$ & 0.0427 & 0.0101 & 0.6293 & 0.2628 & 0.0833 & 0.0129 \\
    SpatBHM 3 & $\mu^0$, $\mu^1\,(V_\mathrm{max})$, $\mu^2\,(V_\mathrm{m})$ & 0.0436 & 0.01 & 0.6331 & 0.2615 & 0.0806 &  0.0122 \\
    \bottomhline
    \end{tabular}
\end{table}

\begin{figure}
    \centering
    \includegraphics[width=\textwidth]{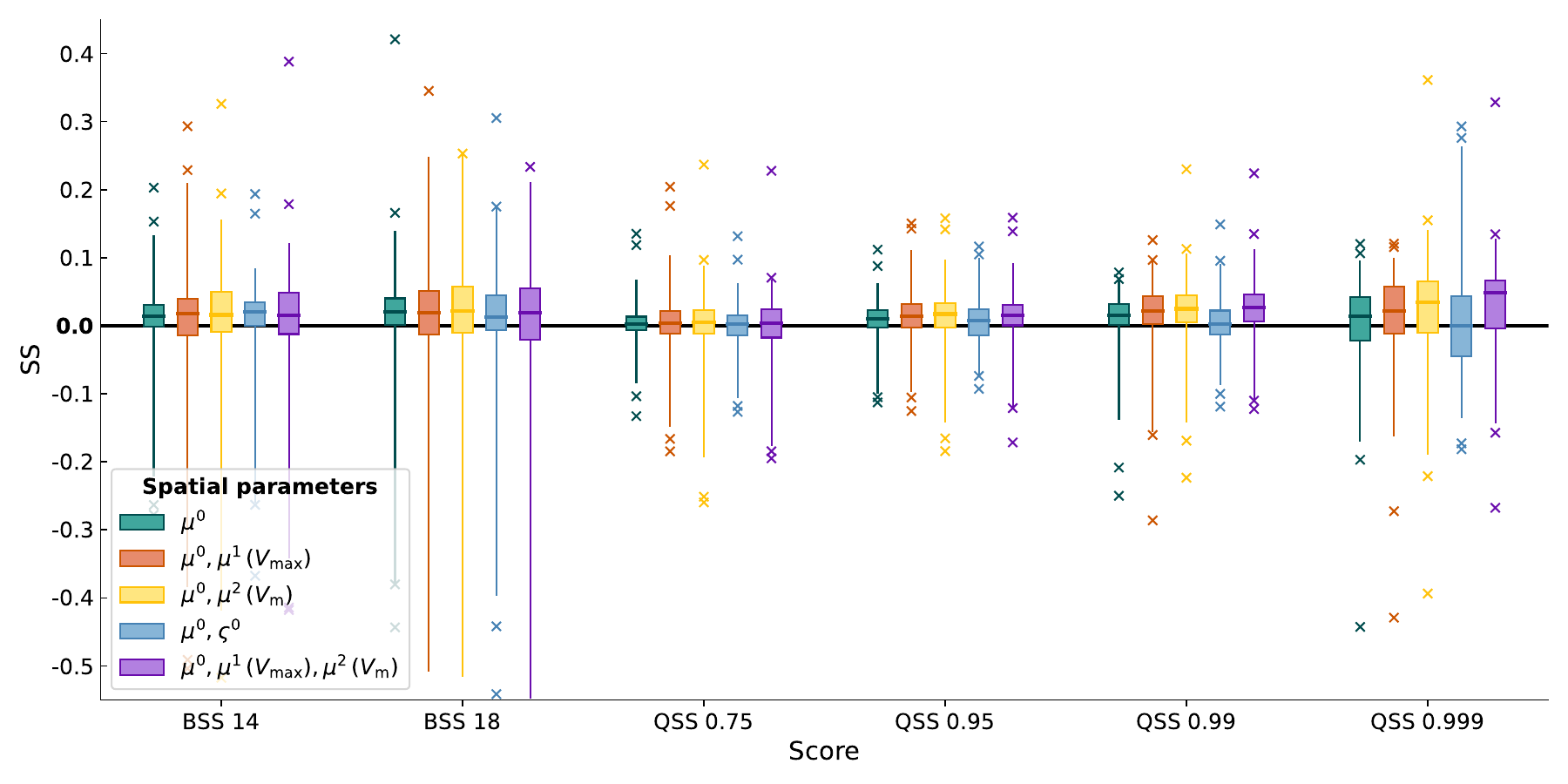}
    \caption{As Fig.~\ref{fig:SkillConstClimLocal} but for cross-validated skill scores of \mbox{SpatBHM} against \mbox{ConstMod}. Model definitions and median scores are given in Table~\ref{tab:ModelOverviewSpatial}.}
    \label{fig:SkillSpatConst}
\end{figure}

The median skill score with respect to \mbox{ConstMod} (Figure~\ref{fig:SkillSpatConst}) is non-negative for all model versions and for all evaluated thresholds and quantiles.
With regard to the threshold excess probabilities, there are not many skill differences among the models which indicates that the gain in benefit from selecting a more complex model is minimal.
For higher quantiles, the skill improvement is higher than for the central region of the predictive distribution, indicating superior performance of \mbox{SpatBHM} for extreme events.

A spatially variable location parameter leads to a significant improvement in the predictive skill of threshold exceedance probabilities and high quantiles.
The inclusion of a spatially variable regression coefficient for the location parameter leads to an improvement in the predictive skill of threshold exceedance probabilities and particularly improves the skill for the extreme quantiles.
A spatially variable regression coefficient for $V_\mathrm{m}$ ($\mu_2$) provides more skill for the upper quantiles than a spatially variable regression coefficient for $V_\mathrm{max}$ ($\mu_1$). 
A spatially variable scale parameter affects different prediction quantities in contradictory ways, rendering its overall effect ambiguous.
It improves the skill for the exceedance probabilities of lower thresholds and reduces the dispersion of the Brier score, but also reduces the skill for all shown predicted quantiles compared to the other spatial model versions.
This is particularly evident for the extreme 0.99 and 0.999 quantiles, where the median skill score compared to \mbox{ConstMod} is approximately zero. 
Whether a spatially variable scale parameter should be selected therefore depends on the desired predicted property.

The addition of a third spatially variable regression coefficient leads to a slight improvement in skill compared to a model with only two spatial parameters. However, this improvement is not as significant and the results for different parts of the distribution are inconsistent.
The third spatial parameter improves the extreme quantiles 0.99 and 0.999, but does not provide much more information for the prediction of threshold exceedances.
Further versions of \mbox{SpatBHM} with varying regression coefficients for the scale parameter (not shown) do not show significantly different results from those described above and are less skillful. We therefore proceed with \mbox{SpatBHM 2b}, while further refinements remain possible.

\begin{table}
    \caption{Parameter estimations for \mbox{SpatBHM} within their 99 \unit{\%} confidence intervals, assuming normal marginal posterior distributions. The values are obtained from training the model on the complete training data set and using Eq. (\ref{eq:AltitudeEuclidean}) as distance metric. Cells with only one number for the spatial parameters indicate spatially constant parameters.}
    \label{tab:SpatbhmParvalues}
    \centering
    \begin{tabular}{cc|cccccc}
    \tophline
    \multicolumn{2}{c|}{\textbf{Parameter}} & \textbf{SpatBHM 1} & \textbf{SpatBHM 2a} & \textbf{SpatBHM 2b} & \textbf{SpatBHM 2c} & \textbf{SpatBHM 3}\\
    \middlehline
    \multirow{3}{*}{ \large{$\mu^0$} }  & $\alpha_{\mu^0}$ & $4.78 \pm 0.2$ & $4.78 \pm 0.24$ & $4.80 \pm 0.27 
    $ & $4.78 \pm 0.22$ & $4.82\pm 0.23$\\
     & $\sigma_{\mu^0}$  & $0.56 \pm 0.11$ & $0.70 \pm 0.16$ & $0.69 \pm 0.14$ & $0.61\pm0.13$ & $0.67 \pm 0.14$\\ 
     & $\rho_{\mu^0}$ & $32 \pm 17.5$ & $37 \pm 18.1$ & $38 \pm 22.2$ & $32 \pm 18.3$ & $30 \pm 18.7 $\\
    \rule{0pt}{4pt} & \rule{0pt}{4pt} & \rule{0pt}{4pt} & \rule{0pt}{4pt} & \rule{0pt}{4pt} & \rule{0pt}{4pt} \\
    \multirow{3}{*}{\large{$\mu^1$} }   & $\alpha_{\mu^1}$ & \multirow{3}{*}{$1.63\pm 0.03$} & $1.56 \pm 0.17$ & \multirow{3}{*}{$1.47 \pm 0.03$} & \multirow{3}{*}{$1.61 \pm 0.03 $} & $1.54\pm0.20$\\
     & $\sigma_{\mu^1}$ & & $0.4 \pm 0.10 $ & & & $0.35 \pm 0.11$\\ 
     & $\rho_{\mu^1}$ & & $ 45 \pm 26.8$ & & & $76 \pm 39.4$\\
    \rule{0pt}{4pt} & \rule{0pt}{4pt} & \rule{0pt}{4pt} & \rule{0pt}{4pt} & \rule{0pt}{4pt} & \rule{0pt}{4pt} \\
    \multirow{3}{*}{ \large{$\mu^2$} } & $\alpha_{\mu^2}$ & \multirow{3}{*}{$-0.57 \pm 0.03$} & \multirow{3}{*}{$-0.37 \pm 0.04 $} & $-0.30 \pm 0.28$ & \multirow{3}{*}{$-0.52 \pm 0.03$} & $-0.34\pm 0.34$\\
     & $\sigma_{\mu^2}$  & & & $0.60 \pm 0.16 $ & & $0.63 \pm 0.22$\\ 
     & $\rho_{\mu^2}$ & & & $62 \pm 38.4$ & & $68 \pm 37.3$\\
    \rule{0pt}{4pt} & \rule{0pt}{4pt} & \rule{0pt}{4pt} & \rule{0pt}{4pt} & \rule{0pt}{4pt} & \rule{0pt}{4pt} \\
    \multicolumn{2}{c|}{\large{$\mu^z$}} & $0.23 \pm 0.08$ & $0.26 \pm 0.10$ & $0.32 \pm 0.10$ & $0.31 \pm 0.08$ & $0.33\pm0.09$\\
    \rule{0pt}{4pt} & \rule{0pt}{4pt} & \rule{0pt}{4pt} & \rule{0pt}{4pt} & \rule{0pt}{4pt} & \rule{0pt}{4pt} \\
    \multirow{3}{*}{ \large{$\varsigma^0$}} & $\alpha_{\varsigma^0}$ & \multirow{3}{*}{$0.491 \pm 0.005$} & \multirow{3}{*}{$0.473 \pm 0.005$} & \multirow{3}{*}{$0.468 \pm 0.005$} & $0.49 \pm0.08$ & \multirow{3}{*}{$0.464 \pm 0.005$}\\
     & $\sigma_{\varsigma^0}$ & & & & $0.20 \pm 0.05$ &\\ 
     & $\rho_{\varsigma^0}$ & & & & $35 \pm 25.7$ &\\
    \rule{0pt}{4pt} & \rule{0pt}{4pt} & \rule{0pt}{4pt} & \rule{0pt}{4pt} & \rule{0pt}{4pt} & \rule{0pt}{4pt} \\
    \multicolumn{2}{c|}{\large{$\varsigma^1$}} & $0.373 \pm 0.009$ & $0.365 \pm 0.009$ & $0.348\pm0.009$ & $0.40 \pm 0.01$ & $0.347\pm 0.009$\\
    \rule{0pt}{4pt} & \rule{0pt}{4pt} & \rule{0pt}{4pt} & \rule{0pt}{4pt} & \rule{0pt}{4pt} & \rule{0pt}{4pt} \\
    \multicolumn{2}{c|}{\large{$\varsigma^2$}} & $-0.224 \pm 0.008$ & $-0.240 \pm 0.008$ & $-0.227\pm0.008$ & $-0.26 \pm 0.01$ & $-0.227\pm0.008$\\
    \rule{0pt}{4pt} & \rule{0pt}{4pt} & \rule{0pt}{4pt} & \rule{0pt}{4pt} & \rule{0pt}{4pt} & \rule{0pt}{4pt} \\
    \multicolumn{2}{c|}{\large{$f_\mathrm{z}$}} & $116 \pm 73.8$ & $157 \pm 69.6$ & $184 \pm 81.2$ & $133 \pm 68.9$ & $194 \pm 78.8$\\
    \bottomhline
    \end{tabular}
\end{table}
\clearpage

\section{GEV shape of wind gust distributions}
\label{A:GEV_shape}
The GEV shape parameter $\xi$ has a great effect on the distribution. E.g., if $\xi<0$, the distribution becomes a Weibull-Type distribution with an upper bound on the supported values, while the Gumbel-type distribution for $\xi=0$ is unbounded. As $\xi$ is difficult to infer stably from the data, we decided to set it to a fixed and spatially constant value.  As we are hesitant to exclude potential extreme events, we would like to intentionally omit the Weibull-type distribution as a valid option, although it may theoretically occur. 

To determine an appropriate choice for the constant $\xi$, we fitted non-stationary GEV-models at each station to obtain local estimates of $\xi$ for the gust model. We calculated the model for 1000 bootstrap iterations in order to obtain 95\%-credibility intervals for the parameter estimation. If the credibility intervals include 0, we conclude there is no signal for $\xi\geq 0$ at the station and the Gumbel-assumption is justified. If the credibility intervals are entirely positive, we say wind gusts at the station follow a Fr\'echet-type distribution. Finally, if the entirety of the credibly interval is below 0, we say that wind gusts follow a Weibull-type distribution, which we would like to exclude. Therefore, if all stations are either Gumbel or Weibull, we regard the assumption that $\xi\equiv 0$ as justified for \mbox{SpatBHM}.

We found only one station with a Fr\'echet-type distribution, but the 95 \%-credibility interval is entirely $\xi<0.1$, so that the value is small. The majority with 84 out of 109 stations show a Weibull-Type distribution with mean shape values between $-0.3<\xi<0$. The remaining stations have no clear signal with respect to $\xi$. We could not detect a spatial pattern for $\xi$.
Some stations with low local GEV shape values coincide with locations where our spatial model exhibits limited skill. However, there are also stations with near-zero local GEV shape values, where \mbox{SpatBHM} still performs poorly. Conversely, several stations with a significantly negative $\xi$ are well captured, suggesting that factors other than the local shape parameter influence model performance.
Hence, assuming $\xi=0$ for \mbox{SpatBHM} appears justified or at least not contradicted by the data.

\section{Positive definiteness of the covariance matrix}
\label{A:CovMatrix}
Applying the Mat\'ern-3/2 covariance function is known to be not positive definite on the sphere, when used with the great-circle distance \citep{Gneiting13}. This results from the restriction of a function defined on $\mathbb{R}^+$ to a finite domain. However, for \mbox{SpatBHM} we are only modeling a limited portion on the sphere with maximum distances of $\mathcal{O}(10^3)$ km, so that the investigated area is almost flat. Nevertheless, we like to work on the real distances, which is why we selected the geat-circle distance. 

In earlier stages, we experimented with an exponential covariance kernel, which is known to be positive definite on the sphere. The exponential kernel is equivalent to the Mat\'ern-1/2-Kernel and therefore the roughest member of the family. However, due to the existence of very short distances between some of the training locations, the exponential kernel produced numerical instabilities. Therefore, we selected a function with a higher smoothness, thereby sacrificing the guarantee of a positive definite matrix.

We ran sensitivity tests to ensure that the covariance matrices in \mbox{SpatBHM} remain positive definite in practice for the limited modeling window over central Europe. To this end we calculated the covariance matrix using the coordinates of the training locations and a wide variety of possible values for $f_z$, $\rho$ and $\sigma$, by sampling from their respective prior distributions (Table \ref{tab:priors}) 10000 times. For each set of parameters, we determined whether the resulting covariance matrix remains positive definite.

As shown in Fig. \ref{fig:CovMatrix}, from all 10000 parameter draws, only 3 combinations resulted in a non-positive definite covariance matrix. These draws had in common that the range parameter was very large at values $\rho > 10^5$ \unit{km}, while the values of the altitude scaling factor $f_z$ and the sill $\sigma$ did not visibly affect the outcome. The largest range parameters occurring in the posterior samples from \mbox{SpatBHM} are of $\mathcal{O}(10^3)$ \unit{km}, so we conclude that the assumption that the covariance matrices in \mbox{SpatBHM} remain positive definite in practice, is justified.

Additionally, we conducted a sensitivity test of \mbox{SpatBHM} using an even smoother Mat\'ern-5/2 covariance function. While the smoother covariance function had little effect on visible roughness of the $\mu^0$-field and the model’s predictive performance, its numerical stability compared to the exponential covariance function supports the use of a higher smoothness assumption for our modeling area.

\begin{figure}
    \centering
    \includegraphics[width=0.5\linewidth]{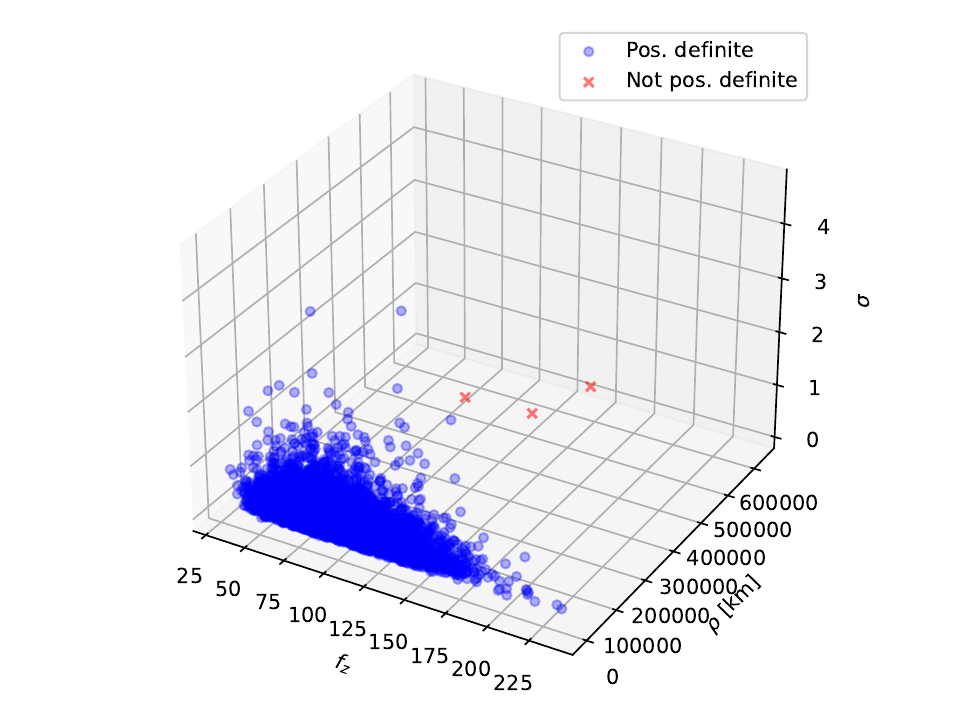}
    \caption{Positive definiteness of the covariance matrix depending on the covariance parameters. }
    \label{fig:CovMatrix}
\end{figure}

\noappendix       




\appendixfigures  

\appendixtables   


\authorcontribution{Petra Friederichs is responsible for conceptualization of this work, acquisition of funding and editing of the manuscript. Philipp Ertz is responsible for conducting the investigation including data processing, model training and prediction, verification and visualization, and for writing the original draft of the manuscript.} 

\competinginterests{The authors declare that they have no conflict of interest.} 


\begin{acknowledgements}

This work has been conducted in the framework of ComingDecade (BMFTR FKZ 01LP2327F) and ClimXtreme II - Modul B (BMFTR FKZ 01LP2323A) both funded by the German Bundesministerium für Forschung, Technologie und Raumfahrt, and the Hans-Ertel-Centre for Weather Research funded by the German Bundesministerium für Verkehr und Digitale Infrastruktur (FKZ 4823DWDP5A). We are grateful to Prof. Dr. Andreas Hense, University of Bonn, for fruitful discussions and constructive suggestions. Finally, we want to thank Sebastian Buschow for reviewing the manuscript and valuable feedback on the accessibility of the accompanying code.

\end{acknowledgements}







\bibliographystyle{copernicus}
\bibliography{article}
\end{nolinenumbers}

\end{document}